# On Yukawa Potential Centrality for Identification of Influential Spreaders in Complex Networks

Pouria Bazyarrezaei and Mohammad Abdollahi Azgomi[*]
Trustworthy Computing Laboratory, School of Computer Engineering,
Iran University of Science and Technology, Tehran, Iran
E-mail: p_bazyarrezaei@alumni.iust.ac.ir and azgomi@iust.ac.ir

**Abstract**

Identifying influential nodes in complex networks is a fundamental challenge for understanding how information, influence, and contagion propagate through interconnected systems. Conventional centrality measures—particularly gravity-based models—often depend on pairwise interaction forces and a fixed radius of influence, which oversimplify the heterogeneous and dynamic nature of real networks. To overcome these limitations, this study proposes a novel non-interactive, action-based model, termed Yukawa Potential Centrality (YPC)**,** which adapts the physical Yukawa potential to the topology of complex networks. Unlike gravity models, YPC computes a scalar potential for each node rather than pairwise forces, dynamically adjusting its radius of influence according to local structural properties. This formulation establishes a physically interpretable bridge between potential theory and network science, while significantly reducing computational complexity—from quadratic to near-linear time. The model is evaluated across both synthetic and real-world social networks, and its node rankings are compared with classical centrality indices and epidemic spreading models (SI and SIS). Experimental findings reveal that YPC exhibits a strong positive correlation with the SIS model and effectively isolates key spreaders, even within highly irregular topologies. These results demonstrate that YPC provides a scalable, adaptive, and theoretically grounded framework for influence analysis in social, biological, and communication networks.

## 1. Introduction

Complex networks consist of a set of nodes and the connections between them, exhibiting properties such as small-world characteristics, scale-freeness, self-organization, and self-similarity. Due to their flexibility and ability to describe real-world systems, these networks have found widespread applications in various domains, including biology, social sciences, and multi-agent systems [1] [2]. Analyzing these networks is crucial not only for understanding their structure but also for examining their interactions and internal dynamics. One of the fundamental challenges in this field is identifying nodes that have the greatest impact on network behavior. These influential nodes play a crucial role in information diffusion, network stability, and resilience to sudden changes [3] [4]. Consequently, accurately identifying these nodes is critical for optimizing and efficiently managing networks [5].

Despite the numerous methods proposed for identifying influential nodes, most approaches primarily focus on the topological features of the network. One of the most common models is the Gravity Model (GM), which estimates the influence of nodes by drawing inspiration from Newton's law of gravitation. However, these models face two key limitations: (1) the assumption of a fixed radius of influence, whereas in real networks, this radius is dynamic and dependent on the network structure; and (2) the neglect of node positions, despite their significant impact on interactions and influence. For instance, nodes located in central regions of a network typically establish more connections than peripheral nodes, yet this distinction is often overlooked in existing models [6].

---

[*] Address for correspondence: School of Computer Engineering, Iran University of Science and Technology, Hengam St., Resalat Sq., Tehran, Iran, Postal Code: 16846-13114, Fax: +98-21-73225322, E-mail: azgomi@iust.ac.ir.

In recent years, identifying influential nodes has become a crucial research topic in network science. These nodes, also referred to as crucial spreaders, play a vital role in network cohesion, information dissemination, epidemic control, and the formation of communication flows. However, accurately identifying them remains an ongoing scientific challenge. The significance of this issue is further amplified by its diverse applications, such as analyzing social networks, preventing disease outbreaks, and optimizing knowledge graphs [7] [3]. Nevertheless, many existing models rely solely on topological features and fail to fully account for real-world node interactions. Therefore, it is essential to develop novel methods that incorporate not only topological analysis but also node positioning and dynamic influence radius [6].

This paper presents an innovative model for identifying influential nodes in complex networks. By leveraging the Yukawa potential, this model introduces an action-based approach for quantifying node influence. Unlike existing methods that assume a fixed influence radius, the proposed model dynamically calculates the radius of influence based on the topological properties of the network. This capability allows the model to analyze different network structures with greater accuracy and improve the identification of key nodes. Furthermore, experimental results indicate that the proposed model outperforms existing methods in detecting influential nodes. The structure of this paper is as follows: Section 2 reviews the theoretical background and fundamental concepts related to complex networks. Section 3 presents a literature review and comparison of existing models. Section 4 discusses the motivations behind this study and key research challenges. Section 5 introduces the Yukawa Potential Centrality (YPC) model and explains its methodology in detail. Section 6 presents experimental evaluations using synthetic and real-world datasets, comparing the performance of YPC with existing methods. Section 7 examines threats to validity, followed by Section 8, which provides conclusions, and Section 9, which outlines directions for future research.

## 2. Background

This section introduces the fundamental concepts of complex networks. First, influential nodes and their importance in network dynamics are explained. Then, various methods and models for identifying key nodes are reviewed, including network topology, epidemic models, and Kendall's tau correlation coefficient. Finally, Yukawa potential is introduced as the basis for the proposed model for identifying influential nodes in this study.

### 2.1. Influential Nodes and Their Importance

In complex networks, nodes are the fundamental components connected through edges. These nodes play different roles in the structure and dynamics of the network. High-centrality nodes are recognized as influential nodes because they hold significant importance in information distribution and network flow. These nodes can affect the behavior of others and play a crucial role in the overall functionality of the network [8] [9].

Identifying these nodes is essential in complex networks. Influential nodes can rapidly affect a large portion of the network, facilitating the spread of information, diseases, rumors, or other network-related phenomena [1]. They are vital in social marketing, epidemic control, and misinformation containment, where their identification can enhance network efficiency [4].

Additionally, in social networks and transportation systems, nodes with higher interaction levels or strategic topological positions tend to have a greater impact, and disruptions in their functionality can lead to widespread consequences for the entire network [10] [11].

## 2.2. Network Topology

Network topology refers to the overall structure of a network and the way nodes are connected. It consists of nodes (vertices) and edges (links between nodes) that describe the interactions and information flow within the network. In complex networks, understanding the topology can help in identifying influential nodes [9].

To analyze the importance of nodes, various centrality measures have been developed, including:

- **Degree Centrality (DC):** Measures the number of direct neighbors of a node. A node with more direct connections has a higher degree centrality [12] [4].
- **Betweenness Centrality (BC):** Evaluates a node's importance in shortest paths between other nodes. A node that appears more frequently in shortest paths has a higher influence on network communication [13] [5].
- **Closeness Centrality (CC):** Measures how close a node is to all other nodes in the network, based on geodesic distances [13] [3].
- **Eigenvector Centrality (EVC):** Assigns importance to a node based on the importance of its neighbors. Nodes connected to highly influential nodes receive a higher score [14] [12] [5].
- **Farness Centrality (FC):** Computes the total distance of a node from all other nodes. A higher value indicates that the node is farther from others in the network [15].
- **PageRank Centrality (PRC):** Determines a node's importance based on incoming links from other nodes, considering both quantity and quality of connections [12].
- **Eccentricity Centrality (EcC):** Measures the greatest distance from a node to any other node. Higher values indicate connections to distant network regions [16].

Additionally, one of the key techniques for analyzing network structures is the K-shell decomposition method. This method helps identify influential nodes by categorizing them based on degree values. Nodes are divided into layers (shells), where each shell consists of nodes with similar degree values [4]. The iterative removal process ensures that nodes with higher degrees remain in deeper shells. This method is particularly useful in social networks and communication systems, as nodes in inner shells typically have greater influence on network dynamics [17] [7].

## 2.3. Epidemic Spread Models

Epidemic spread models in complex networks are used to simulate the spread of diseases in populations and social networks. These models leverage network theory and mathematical principles to simulate interactions among nodes and analyze how diseases propagate over time and space. Influential nodes play a key role in either accelerating or controlling the spread of infections. In these models, nodes can represent individuals, locations, or systems, while edges indicate interactions, information exchange, or disease transmission. Nodes with higher connectivity or strategic positioning in the network tend to facilitate faster disease spread.

Several epidemic models have been developed, with the most significant ones described below:

- **The SI Model:** In this model, the population is divided into two categories: Susceptible (S) and Infected (I). Once a node becomes infected, it remains in that state permanently. This model is particularly suitable for chronic or permanent infections [13].
- **The SIR Model:** Introduced by Anderson and May, this model categorizes individuals into Susceptible (S), Infected (I), and Recovered (R) groups. It is appropriate for diseases that grant permanent immunity after recovery, such as measles [18].
- **The SIS Model:** The SIS model applies to diseases where recovered individuals do not acquire permanent immunity and can become susceptible again. This model is particularly useful for modeling seasonal viruses like the common cold [19].
- **The SEIR Model:** An extension of the SIR model, the SEIR model introduces an Exposed (E) state, representing individuals who have been infected but are not yet contagious. This model is particularly relevant for diseases with an incubation period, such as COVID-19 [20].

A key concept in epidemic models is the basic reproduction number $R_0$, which indicates the average number of secondary infections generated by an infected individual. If $R_0 > 1$, the disease spreads rapidly within the population; if $R_0 < 1$, the disease is gradually contained [21]. Studies by Barabási [22] and Kitsak [7] have shown that nodes with high centrality play a significant role in disease propagation speed, and identifying these influential nodes can greatly improve epidemic containment strategies.

## 2.4. Kendall's Tau Correlation Coefficient

Kendall's tau correlation coefficient is a rank-based measure used to assess correlation between two ranked datasets. It was introduced by Maurice Kendall in 1938 and is particularly useful for ordinal data [23]. Kendall's tau is well-suited for comparing ranked or discrete data but is less applicable to continuous datasets.

This coefficient quantifies the agreement or disagreement between two ranked datasets. A value of $\tau = 1$ indicates perfect positive correlation, meaning that an increase in one variable corresponds to an increase in the other. Conversely, $\tau = -1$ signifies perfect negative correlation, implying that an increase in one variable is associated with a decrease in the other. A value of $\tau = 0$ represents no correlation between the variables [15].

To compute Kendall's tau, the data must first be ranked. Then, the number of concordant pairs ($C$) (where ranks are in the same order in both datasets) and discordant pairs ($D$) (where ranks are in the opposite order) are determined. The coefficient is calculated using the formula:

$$\tau = \frac{(C - D)}{\frac{n(n-1)}{2}} \quad (1)$$

where $C$ represents the number of concordant pairs, $D$ the number of discordant pairs, and $n$ the number of observations [23]. This equation defines Kendall's tau as the difference between the number of concordant and discordant pairs, normalized by the total possible pairs in the dataset. Compared to other rank-based correlation coefficients, such as Spearman's correlation, Kendall's tau is more sensitive to disordered rankings, making it more accurate for discrete ordinal data. While Spearman's coefficient relies on squared rank differences, Kendall's tau is based on pairwise agreement counting, making it particularly useful in complex network analysis, social sciences, and medical research.

For instance, in social network analysis, Kendall's tau helps researchers compare different ranking methods for users based on network centrality measures, allowing for a more detailed analysis of influential nodes and their interdependencies [15].

## 2.5. Yukawa Potential

The Yukawa potential is a fundamental model used to describe short-range interactions in physics. It was first introduced by Hideki Yukawa in 1935 to explain the interactions between nucleons within the atomic nucleus. In this model, attractive or repulsive forces between particles are mediated by the exchange of bosons, such as mesons. The mathematical formulation of the Yukawa potential is given by:

$$V_{Yukawa}(r) = -g^2 \frac{e^{-\alpha m r}}{r} \quad (2)$$

where, $g$ is the coupling constant, $m$ represents the mass of the mediating boson, $\alpha$ is a scaling constant, and $r$ denotes the distance between two interacting particles.

Due to the exponential decay term $e^{-\alpha m r}$, the Yukawa potential diminishes rapidly over long distances, characterizing its short-range nature. The radius of influence in the Yukawa model refers to the distance at which the potential significantly decreases and is determined by the parameter $\mu$ defined as $\mu = \alpha m$. A larger $\mu$ results in a shorter interaction range, meaning that the force is only significant at short distances. This radius is approximately given by $\frac{1}{\mu}$ and, in nuclear physics, it represents the range of nuclear forces and particle interactions. In nuclear physics, the Yukawa potential is employed to model short-range interactions between nucleons, with mesons acting as mediators of these interactions. This model has contributed to explaining the attractive forces between protons and neutrons as well as the stability of heavy nuclei [24].

Additionally, the Yukawa potential has applications in various other domains, such as plasma physics and inter-particle systems. In electrical plasma, it is used to model interactions between charged particles, as it effectively describes the rapid decay of electrostatic forces with increasing distance [25] [26].

This potential also plays a role in modified gravity theories, where models like the Yukawa potential are used to refine gravitational interactions across different scales, including galactic and cosmological distances. Recent studies suggest that combining the Yukawa potential with a finite influence radius can lead to more comprehensive models for describing inter-particle forces and natural phenomena [24].

Furthermore, the Yukawa potential is utilized in modeling the behavior of complex and social systems. In this field, it serves to simulate weak, short-range interactions between various elements. Specifically, in social and economic networks, this model is applied to simulate information propagation, social influence effects, and analyze complex interactions within networks [26].

### Relation to the Coulomb Potential

If a mediating particle has zero mass (i.e., $m = 0$), the Yukawa potential reduces to the Coulomb potential, which has an infinite range:

$$m = 0 \Rightarrow e^{-\alpha m r} = e^0 = 1 \quad (3)$$

Thus, Equation (2) simplifies to the Coulomb potential:

$$V_{Coulomb}(r) = -g^2 \frac{1}{r} \qquad (4)$$

where the scaling constant is given by:

$$g^2 = \frac{q}{4\pi\varepsilon_0} \qquad (5)$$

where $q$ represents the electric charge and $\varepsilon_0$ is the permittivity of free space.

Equation (4) can be further transformed into an interaction equation for the Coulomb force:

$$\vec{E} = -\vec{\nabla}V \xrightarrow{\vec{E}=\frac{\vec{F}}{q_0}} \vec{F} = \frac{qq_0}{4\pi\varepsilon_0 r^2}\hat{r} \qquad (6)$$

where $q$ and $q_0$ are the charges of interacting particles, $\vec{E}$ represents the electric field, $\vec{F}$ is the Coulomb force, and $\hat{r}$ is the unit vector in the direction of interaction.

Figure (1) illustrates the long-range behavior of Yukawa and Coulomb potentials when $g = 1$. As shown, the Coulomb potential remains significant at large distances, whereas the Yukawa potential decays rapidly to zero. However, both potentials maintain nonzero values even at extended ranges.

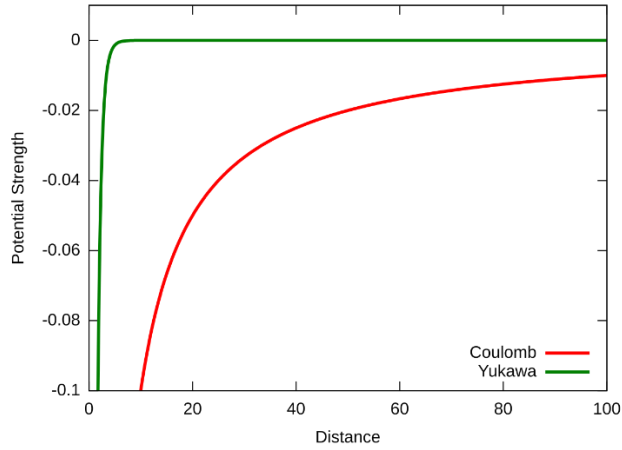

**Figure 1.** Long-range comparison of Yukawa and Coulomb potentials for $g = 1$ g=1 [27].

## 3. Related Work

Identifying influential nodes in complex networks is a fundamental topic in network analysis. Many existing methods for assessing node influence rely on network centrality models; however, gravity-based models offer a distinct approach by incorporating both influence radius and spatial positioning of nodes. This section examines the gravity model and its subsequent developments.

### 3.1. Gravity Model for Identifying Influential Nodes

The gravity model is one of the most widely used methods for modeling interactions in spatial systems and complex networks. Also referred to as gravity centrality, this model was introduced by Ma Ling-ling et al. and is inspired by Newton's law of gravitation, treating interactions between nodes similarly to the gravitational force between two masses. In general, the gravity model predicts interactions between two elements based on two key variables: (1) Intrinsic attributes of the elements (e.g., population in geographical systems or connectivity in social networks). (2) Distance between them. This model suggests that interaction strength between two elements increases with their intrinsic attributes but decreases with increasing distance. The relationship is formulated as follows:

$$I_{ij} = k \frac{P_i . P_j}{D_{ij}^b} \qquad (7)$$

where $I_{ij}$ represents the interaction intensity between elements $i$ and $j$, $P_i$ and $P_j$ denote their intrinsic properties (such as node degrees in complex networks), $D_{ij}$ is the distance between them, $b$ is the distance exponent, and $k$ is a scaling constant. As observed, Equation (7) is analogous to Equation (6), exhibiting an inverse-square relationship. Similar to Newtonian gravity, this model captures the inverse-square relationship of distance effects on interactions in networks [6] [28].

## 3.2. Extensions of the Gravity Model Considering Variable Influence Radius

Despite the success of the gravity model in identifying influential nodes, one of its primary limitations is the assumption of a fixed influence radius for nodes. However, in real-world networks, the impact of nodes is influenced by various factors, such as spatial position and network topology. Consequently, several studies have aimed to address this limitation.

Shuyu Li et al. proposed a modified gravity model in which the influence radius of nodes was treated as a variable. Their model demonstrated that more central nodes in a network tend to have a larger influence radius and stronger interactions with other nodes. This approach significantly improved the accuracy of identifying key nodes and showed a stronger correlation with network dynamics [13].

Similarly, Guiqiong Xu introduced a refined adaptation of the gravity model that, in addition to a dynamic influence radius, also incorporated the effect of network topology. This model revealed that influential nodes can play different roles in information dissemination or flow regulation, depending on their spatial position. Notably, in scale-free networks, the impact of central nodes was found to be significantly higher than what the classical gravity model predicts [1].

## 3.3. Model Comparison

The gravity model and its extensions provide a framework for evaluating node influence. However, the assumption of a fixed influence radius in the classical gravity model presents a limitation that does not hold in many real-world networks. One advantage of the classical model is its simplicity and efficiency in homogeneous networks, but its accuracy declines in heterogeneous networks. In contrast, more advanced models that incorporate a dynamic influence radius have demonstrated greater accuracy in describing the behavior of influential nodes. Nevertheless, these models still require further optimization, as determining an appropriate criterion for adjusting the influence radius remains a significant challenge. The effective gravity model (EGM) offers higher accuracy than the classical gravity model because it can better simulate network dispersion and dynamics. However, its computational complexity is higher, which can make processing large networks more time-consuming. The communicability-based adaptive gravity model (CAGM) achieves even greater accuracy by incorporating more detailed information about the network structure and the role of inter-node connections. However, its higher computational cost and greater demand for precise data remain key challenges.

In this study, the Yukawa Potential Centrality (YPC) model is proposed as a novel approach to address these limitations. Unlike gravity-based models, which rely on direct interactions between nodes, YPC leverages the Yukawa potential to dynamically compute the influence of each node without assuming a fixed influence radius. This characteristic makes YPC a flexible and precise model for analyzing complex networks.

## 4. Motivations and Objectives

One of the fundamental challenges in complex network analysis is the precise identification of influential nodes. These nodes play a crucial role in information dissemination, epidemic control, network stability, and various social and biological phenomena. Several methods have been proposed for identifying these nodes, yet most either rely on topological metrics or use complex interaction-based models to compute node influence. Among these, gravity-based models, which operate based on the gravitational force between nodes, have emerged as a prominent approach. However, such models typically require computationally expensive calculations to determine pairwise interactions, leading to high computational costs.

The proposed Yukawa Potential Centrality (YPC) model, unlike interaction-based methods, is an action-based model. Instead of computing forces between nodes, YPC directly evaluates the potential of each node. This feature allows YPC to measure node influence independently, without relying on pairwise interactions, significantly reducing computational costs. In interaction-based models, such as gravity models, the computational process becomes highly expensive and complex, particularly in large-scale networks. In contrast, YPC utilizes potential as a measure of influence, which is computationally simpler than pairwise interactions.

Another primary motivation of this research is to overcome the limitations of gravity-based models. In many classical gravity models, the influence radius of nodes is assumed to be fixed, whereas in real-world networks, a node's impact is dynamic and depends on the network structure. YPC addresses this issue by leveraging the exponential nature of the Yukawa potential, which enables the influence radius to be dynamically adjusted based on the network topology. This feature makes YPC a flexible and accurate model for identifying influential nodes, making it suitable for various applications, including social networks, communication systems, and biological networks.

Key objectives of this study are as follows:
1. Developing an action-based model for identifying influential nodes without requiring pairwise interaction computations.
2. Reducing computational costs in key node identification by using potential equations instead of force-based calculations.
3. Establishing a flexible method for dynamically adjusting the influence radius of nodes based on network structure.
4. Evaluating YPC's performance against centrality measures and assessing its accuracy and efficiency in complex network analysis.

## 5. Proposed Model

The analysis of topological features in networks and the application of scientific models in complex networks have led to significant advancements in identifying influential nodes. Most existing methods rely on network topological metrics; researchers typically map these topological features onto established scientific models such as the gravity model, Poisson distribution, and probabilistic diffusion models, subsequently developing their proposed models based on these frameworks. However, there is still no universally accepted standard for identifying influential nodes in complex networks, and the results of different methods remain dependent on varied assumptions and network structures [4].

Moreover, merely introducing a mathematical model is insufficient for understanding the structure and dynamics of complex networks. A proposed model must align with real-world phenomena and effectively capture actual network behaviors. The diversity of network structures makes empirical validation using real-world data essential. In this study, social networks and message propagation analysis are used as a case study to evaluate and compare the proposed Yukawa Potential Centrality (YPC) model. Social networks provide a suitable platform for analyzing node influence, as message dissemination and node reachability serve as valuable indicators of influence. Accordingly, the primary approach for evaluating YPC is to measure message spread within a social network. Each node's influence is assessed based on its ability to propagate messages and its access to broader network regions. This practical framework facilitates YPC's validation in real-world settings, allowing for a comparative analysis of its accuracy and efficiency against existing methods.

This section introduces the proposed model for identifying influential nodes in complex networks. Built upon Yukawa potential, the model aims to quantify the influence and spread of each node across the network without requiring direct pairwise interactions.

A conceptual overview of the YPC framework is illustrated in Figure (2). The model takes the **adjacency matrix** of the network as input, maps **topological parameters** such as node degrees and neighborhood structures to the Yukawa potential parameters (e.g., coupling constant $g$ and mass $m$), and finally computes a **scalar influence potential** for each node. This process allows YPC to quantify node influence without relying on pairwise interactions.

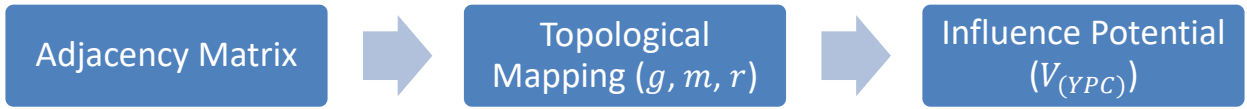

**Figure 2.** Conceptual flow of the Yukawa Potential Centrality (YPC): from network topology (adjacency matrix) to mapped physical parameters and resulting influence potential score.

## 5.1. Description of the Yukawa Potential Centrality (YPC) Model

Before introducing the proposed model, it is essential to define its assumptions, which are as follows:

- **Assumption 1:** Unlike interaction-based models, which explicitly define the relationship between two nodes as an interactive force, YPC computes the overall potential of each node (rather than pairwise interactions). This allows for the identification of key influential nodes in the network. Thus, YPC is an action-based model rather than an interaction-based model.
- **Assumption 2:** For the sake of simplification in initial evaluations, the network graphs considered in this study are undirected and unweighted. A real-world example of such a structure is LinkedIn, which can be modeled as an undirected social network.
- **Assumption 3:** Unlike most existing models for identifying influential nodes (such as various gravity-based models), YPC does not assume that connecting to a highly influential node (e.g., a high-degree node) inherently increases a node's influence. That is, Newton's third law is not incorporated into the model, meaning that in a network, an action does not necessarily result in a reciprocal reaction. This assumption aligns with realistic behaviors observed in many complex networks. For instance, in a social network, if user x sends a message to an influencer y (who has a massive number of followers), the likelihood of user y noticing or responding to user x's message is extremely low due to the high volume of messages received. Therefore, simply connecting to an influential node like y does not guarantee higher influence or message spreadability for user x. Consequently, this assumption reinforces the action-based nature of the YPC model.

**Mapping of Quantities and Modeling:**
Based on the aforementioned assumptions, the Yukawa Potential Centrality (YPC) model is formulated by mapping certain topological parameters of the network to the quantities defined in the Yukawa potential model. This mapping enables the quantification of influence and reachability of each node within the network through the Yukawa potential framework. The following section provides a detailed explanation of each mapped quantity, allowing for a more precise structural analysis of the network within this model.

- **Coupling Constant (Scaling Constant)**
  According to Equation (2), the coupling constant is expressed as $(-g^2)$ to represent a strong attractive force: the negative sign ensures that the effect of the potential is directed toward the influencing particle. This quantity characterizes the "intensity" or "strength" of the potential using a quadratic dependence. In physics, particularly in quantum field theory, the coupling constant serves as a measure of the effective force between particles, modeling a nonlinear (interactive) relationship between the potential sources and their effects.
  
  In the YPC model, the negative sign is negligible. Since the network graph is assumed to be undirected, no explicit definition for directionality is established. However, if a zero potential reference level is assumed and positive potential differences are required in calculations, the negative coupling constant may be useful. Additionally, unlike gravitational or Coulombic potentials, the coupling constant in YPC is treated as a linear term (without a quadratic dependence). This ensures that the potential intensity is directly proportional to the intrinsic strength of each node and causes nodes with higher influence to have stronger effects on their surroundings, and as a result, the potential generated by these nodes has a greater impact on the network than weaker nodes.
  
  According to mapping relation (8), in the YPC framework, the degree of each node is mapped to the coupling constant, determining the intensity of the potential it generates. Consequently, higher-degree nodes produce stronger potentials and exert a greater influence on their surroundings. This feature naturally models the overall impact of a node within the network.
  
  $$-g^2 \xrightarrow{YPC} k \qquad (8)$$

- **Mass**
  Mapping topological parameters to the quantity $\mu = \alpha m$ from Equation (2) is somewhat complex. For simplification, the scaling constant $\alpha$ can be set to 1, so that $\mu$ is equivalent to the particle mass $m$.
  
  In Figure (3), assuming $g = 1$, the variation of potential with increasing distance for different values of $m$ is illustrated. As previously noted, when $m = 0$, the Yukawa potential reduces to the Coulomb potential, meaning the range of influence is theoretically infinite. However, as $m$ increases, the effective range of the particle decreases. If $m$ becomes sufficiently large, the interaction becomes extremely short-ranged, with a rapid decay of potential over very short distances.

By defining a threshold constant $\zeta^\dagger$, the potential decay can be approximated as follows:
$$\lim_{V \to \zeta} dV/dr = 0 \tag{9}$$
This implies that once the potential intensity drops below a threshold value, further increases in distance have no significant effect on the potential.

With these interpretations, to complete the YPC model, it is necessary to map the topological parameter to $m$ so that with its increase, considering the nature of equation (2), not only does the node's influence (i.e., its potential) tend strongly to zero, but this tendency also occurs in a short range. In other words, in the YPC model, for a large value of the topological property such as m for each node, the radius of influence of that node will not be large (deep); just like the abduction of nucleons in the very limited and small space of the atomic nucleus (compared to the dimensions of the atom itself).

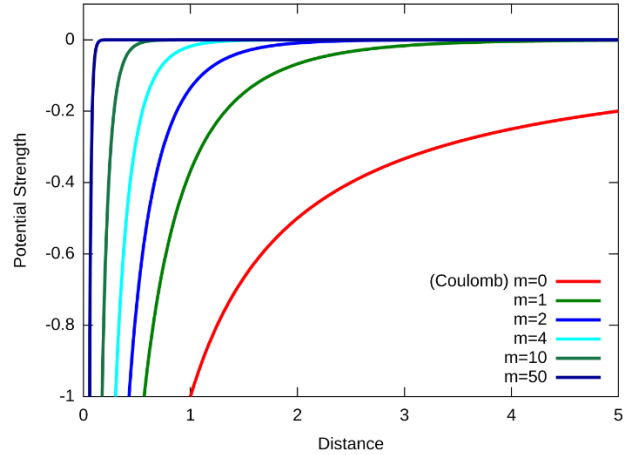

**Figure 3.** Comparison of Yukawa potentials for different values of $m$ where $g = 1$ [27].

At first glance, considering Assumption 3 in Section 5.1, the sum of the degrees of the neighboring nodes of the initiator node can be mapped as the topological parameter corresponding to $m$. This is because each neighbor within the effective radius is assumed to ignore the received influence with a certain probability.

For instance, in the case study approach, if the initiator node sends a message to a neighboring node with high degree centrality, the probability of message propagation by this neighboring node—and consequently, the influence radius—decreases significantly. Based on this explanation, in an undirected network $G = \langle V, E \rangle$, using Equation (10) and mapping relation (11), the YPC equation is expressed as Equation (12):
$$\Gamma_r(o) = \{n_r | n_r \in V(o, n) \in E\} \tag{10}$$
where $\Gamma_r(o)$ is the set of neighbors of the initiator node $o$ within radius r, and $n_r$ represents a neighbor of $o$ within $\Gamma_r(o)$.

---

$^\dagger$ $\zeta$ is considered a negative number between $-1$ and $0$ and very close to zero.

$$m \xrightarrow{YPC} \sum_{n \in \Gamma_r(o)} k_n \qquad (11)$$

$$V_{YPC}o(r) = -k_o \frac{e^{-\sum_{n \in \Gamma_r(o)} k_n r}}{r} \qquad (12)$$

where $k_o$ is the degree of the initiator node $o$, and $k_n$ is the degree of node $n$ within radius $r$ from $o$.

By incorporating $\zeta$, a maximum threshold for influence radius $r$ can be considered. This means that neighbors beyond this threshold will no longer contribute to the potential intensity of node $o$.

Two important points regarding Equation (12) should be noted:

1. Just as $m$ in the original Yukawa potential equation, Equation (2), represents the mass of the mediator particle responsible for force transmission, the term $\sum_{n \in \Gamma_r(o)} k_n$ in the YPC equation denotes the magnitude of topological parameters of the elements (i.e., neighboring nodes) that mediate the effect propagation. This conceptual similarity reinforces the significance of the performed mapping.
2. The interpretation of this mapping suggests that, although the very nature and structure of a network inherently facilitate influence propagation (message transmission), the network's topology and node arrangement can simultaneously act as a mesh (filter), impeding this propagation. This phenomenon is defined as **self-inhibition** of the network.

To further examine this mapping, we can use boundary value analysis. Figure (4) illustrates two nodes: one with an infinitely high degree and another with the minimum possible degree ($k = 1$). In case (a), the blue node, which is connected only to the red node, initiates a message. The probability that the red node ignores the received message is infinitely high, limiting the influence radius to a single edge. However, in case (b), if the red node sends a message to the blue node, the blue node will receive it with a probability of one. If it has access to a limited number of other nodes, it may contribute to extending the influence radius and spreading the red node's message.

Upon closer inspection, it becomes evident that the current mapping has deficiencies in accurately reflecting node influence in the network. As shown in Figure (5), the yellow node, which has a significantly higher degree (on the same order as the red node), sends a message to the red node, while the blue node, which has only one neighbor, also sends a message to the red node. However, based on the previous explanations, both messages are ignored by the red node and are not transmitted. This contradicts real-world node behavior in complex networks. Even though the yellow node, compared to the blue node, has a much higher degree and, according to the initial mapping, should have a greater influence, its message propagation is blocked in the same manner as the blue node's message. In other words, the red node does not differentiate between the yellow node and the blue node, effectively nullifying the influence factor $k_o$ in Equation (12) for impact within the influence radius.

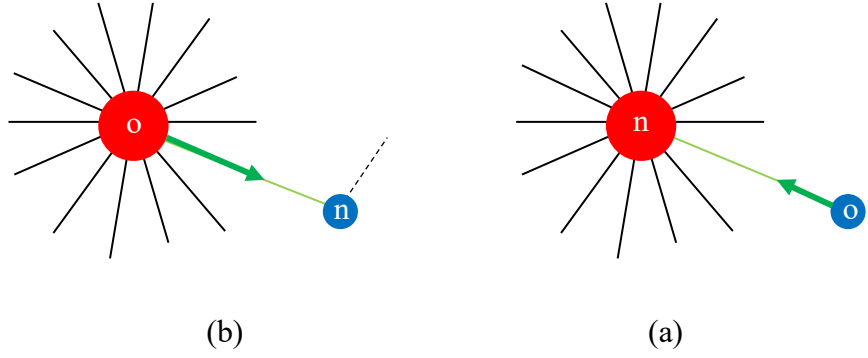

**Figure 4.** Two nodes, one with the minimum degree and the other with the maximum degree: (a) The blue node is the initiator, and the red node is its neighbor. (b) The red node is the initiator, and the blue node is its neighbor.

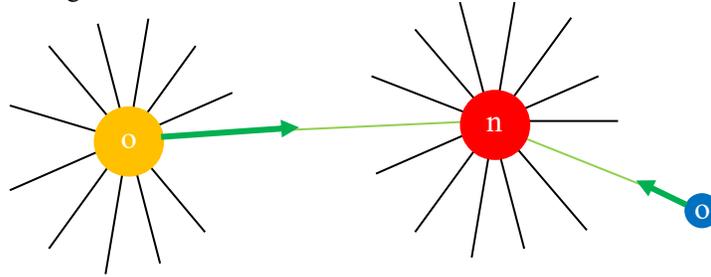

**Figure 5.** Both the yellow and blue nodes send messages to the red node, but both messages are ignored.

For example, in social networks, if two influencers are connected (red and yellow nodes in Figure (5)), the network structure ensures that they treat each other differently compared to regular users. The second mapping fails to satisfy this requirement.

To address the deficiency in mapping to $m$, the mapping relation (11) must be updated to relation (13), and Equation (12) must be modified to Equation (14):

$$m \xrightarrow{YPC} \frac{\langle k_n \rangle}{k_o} \tag{13}$$

$$V_{YPC}o(r) = -k_o \frac{e^{-\frac{\langle k_n \rangle}{k_o}r}}{r} \tag{14}$$

where $k_o$ is the degree of the initiator node $o$, and $\langle k_n \rangle$ is the average degree of the nodes in the set $\Gamma_r(o)$.

The reason for using the averaging operator in Equation (14) is to bring the degrees of the neighboring nodes within the effective radius $r$ of the initiator node $o$ numerically closer to $k_o$. It is as if node $o$ is interacting with a single node of degree $\langle k_n \rangle$ rather than a set of neighbors with varying degrees. Thus, in the example of Figure (5), considering Equation (14), the average degree of the yellow node's neighbors is adjusted relative to the yellow node's degree (a large number). Consequently, its influence radius is no longer as short as that of the blue node, which remains limited to a single edge. Based on this, Equation (14) for the yellow node takes the form of Equation (15), while for the blue node, it takes the form of Equation (16):

$$\lim_{\langle k_n \rangle \to k_o} \frac{\langle k_n \rangle}{k_o} = 1 \to V_{YELLOW} o(r) = -k_o \frac{e^{-r}}{r} < \zeta \tag{15}$$

$$\lim_{\langle k_n \rangle \to \infty} \frac{\langle k_n \rangle}{k_o} = \infty \to V_{BLUE} o(r) = -k_o \frac{e^{-\infty}}{r} = 0 > \zeta \Rightarrow r = 1 \tag{16}$$

Thus, in Equation (14), the factor $\frac{\langle k_n \rangle}{k_o}$ serves as a prioritization mechanism.

### 5.2. Algorithm of the YPC Model

The flowchart of the YPC model is shown in Figure (6).

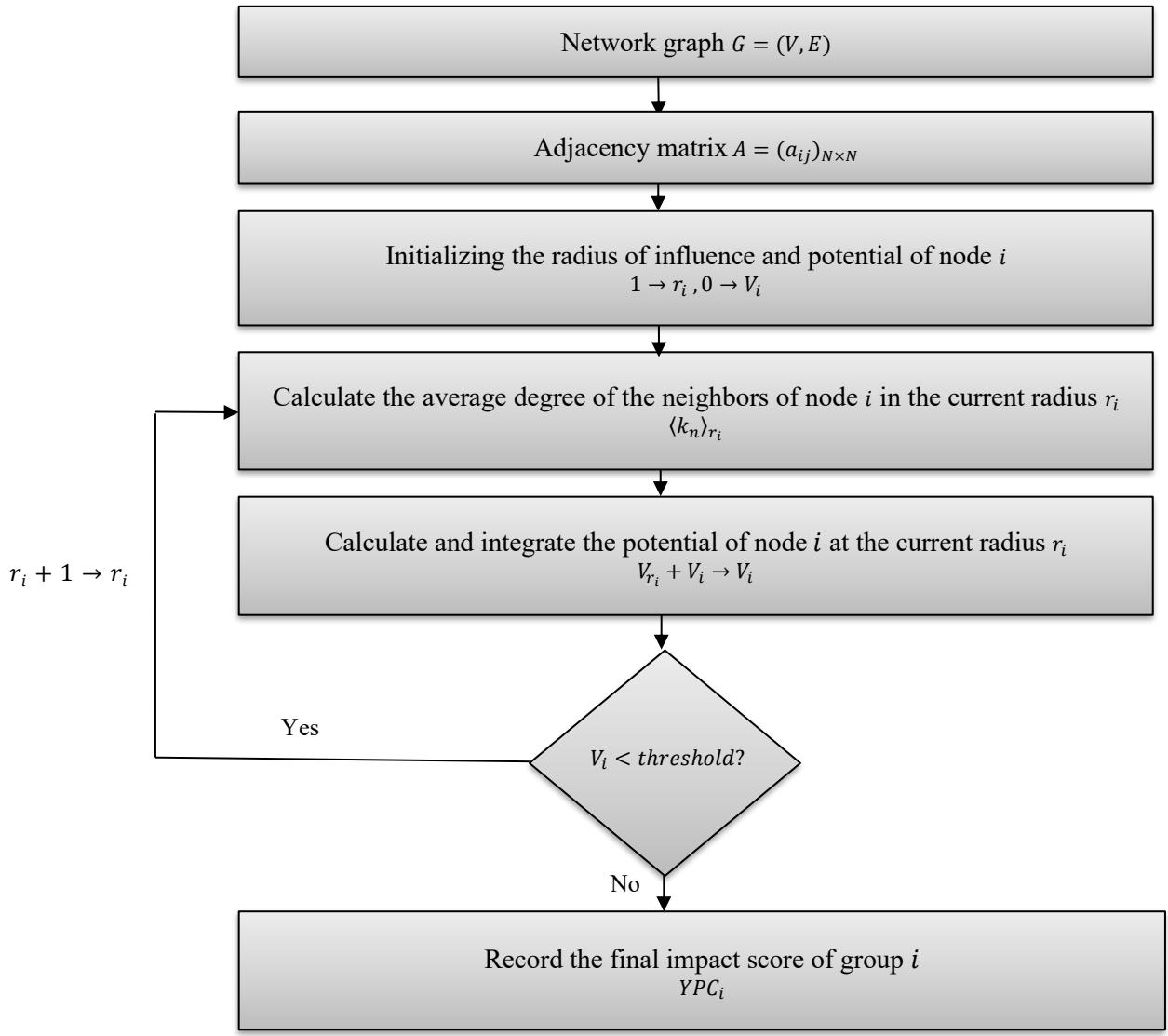

**Figure 6.** Flowchart of the YPC algorithm

In physics, potential energy is interpreted as the configuration energy of a mechanical system, where the relative positioning of the system's elements affects the final energy value. Inspired by this interpretation, if the stopping condition is not met at any step of the calculation ($V_i <$ $threshold$), then, with each increment in the influence radius of the initiator node and the neighborhood order, the newly computed potential at the next neighborhood level is aggregated with the potentials from previous steps according to the superposition principle. This ensures that the configuration of nodes in proximity to each other is taken into account. Once the stopping condition is satisfied, the sum of the computed potentials is considered the final influence score of the initiator node.

The YPC algorithm is presented in Figure (7). The approximate zero threshold, $\zeta$ is set to $10^{-100}$.

---

*The YCP Algorithm:*
- ***Input:*** **Graph** *file (txtFile), Parameters: α, threshold, maxRadius*
- ***Output:*** *Sorted YPC Scores for all nodes in the graph*
1. *Load the graph G from the input file using Snap library.*
2. ***Initialize****:*
   - *numNodes ← **Total** number of nodes in G*
   - *nodeIds ← **List** of all node IDs*
   - *degrees ← Dictionary of degrees for each node in G*
3. ***For*** *each node i in nodeIds:*
   a. ***Initialize****:*
      - *r ← 1*
      - *currentNeighbors ← **Set** of immediate neighbors of node i*
      - *allNeighbors ← **Set** of all visited neighbors of node i*
      - *impactScores ← **Empty** list*
   b. ***While True****:*
      i. *Calculate:*
         - *avgDegree ← Average degree of all nodes in allNeighbors*
         - *ithNodeDegree ← Degree of node i*
         - *ratio ← avgDegree / ithNodeDegree (if ithNodeDegree > 0; else ratio ← 0)*
         - *impactScore ← $\left(-\frac{ithNodeDegree \cdot e^{-\alpha \cdot ratio \cdot r}}{r}\right)$*
      ii. ***Append*** *impactScore to impactScores*
      iii. *Identify nextNeighbors:*
         - ***For*** *each neighbor in currentNeighbors:*
            - *Add second-order neighbors to nextNeighbors (excluding visited nodes and node i)*
      iv. ***Check*** *termination conditions:*
         - ***If*** *impactScore < threshold OR r ≥ maxRadius OR nextNeighbors is empty:*
            ***Break*** *the loop.*
         - *Else:*
            ***Update****:*
               - *currentNeighbors ← nextNeighbors*
               - *allNeighbors ← allNeighbors ∪ nextNeighbors*
               - *r ← r + 1*
   c. *Calculate:*
      - *totalImpactScore ← **Sum** of all values in impactScores*
   d. ***Save*** *result:*
      - *results.append({"nodeId": i, "YPCScore": totalImpactScore, "radius": r})*
4. ***Sort*** *results based on YPCScore in ascending order.*
5. ***Save*** *sorted results to an output file.*
*End.*

**Figure 7.**     The YPC algorithm

Unlike gravitational centrality models, which compute pairwise forces between all node pairs and therefore exhibit a computational complexity of $O(V^2)$, the proposed YPC model evaluates only scalar potentials for individual nodes.

This non-interactive formulation eliminates the need for pairwise force calculations, effectively reducing the complexity to approximately $O(V.(V + E))$.

Consequently, YPC achieves significantly better scalability on large real-world networks without compromising analytical precision.

The overall time complexity of the algorithm is $O(V.(V + E))$., where $V$ denotes the number of nodes and $E$ denotes the number of edges in the network. Notably, due to the exponential decay property of the Yukawa potential, the influence of each node is limited to a few neighboring layers. This feature ensures that, in practice, the algorithm processes only a small subset of the network, leading to a reasonable execution time even for large-scale graphs.

## 6. Evaluation of the Proposed Model

This section is dedicated to evaluating the Yukawa Potential Centrality (YPC) model and analyzing its results. The performance of YPC is compared with the epidemic models SI and SIS, which serve as evaluation benchmarks, as well as other centrality measures. The objective of this section is to examine the correlation and ranking differences between these models and indices, providing scientific insights into the behavior of nodes in complex networks. By presenting detailed data analyses and empirical results, this section establishes a foundation for the practical application of YPC and its further development.

### 6.1. Comparison Models

To evaluate the YPC model, the simplest epidemic model, SI, was initially used. According to the case study approach, all nodes in a social network are in a susceptible state, meaning they are ready to receive messages. Once a message is received from the initiator node, the state of these nodes changes to infected (or informed). Since the SI model assumes that infected nodes remain in this state permanently, with no possibility of returning to the susceptible or recovered states, it does not allow for repeated interactions among nodes, leading to a saturation effect. This limitation makes SI unsuitable for evaluating the proposed YPC model.

It was found that through trial and error, the SI model's parameters can be adjusted based on the network's topological properties to achieve a strong positive correlation between the rankings produced by SI and YPC using Kendall's tau correlation coefficient. However, due to the extreme simplicity of the SI model and the complexity of the studied network topology, setting SI parameters to achieve this strong correlation introduces an unnatural interpretation: To obtain this correlation, the transmission rate $\beta$ in the SI model must have an inverse relationship with the initiator node's degree $k_o$ and a direct relationship with $k_n$, whereas in YPC, the initiator node's degree has a direct relationship with its influence score.

Another major limitation of the SI model is the occurrence of saturation, meaning that once a node becomes infected (informed), it can no longer be infected again through another path. Since the node is removed from the set of susceptible nodes, this creates an unfair allocation of influence scores to the initiator node for new paths. As a result, the SI model leads to the aforementioned unnatural interpretation and is not suitable for evaluating YPC. Figure (8) illustrates the saturation effect in the SI model.

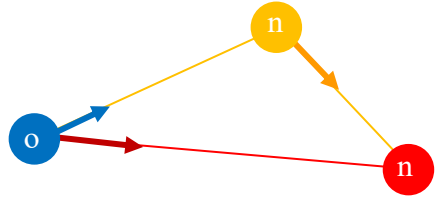

**Figure 8.**      Saturation effect in SI model.

On the other hand, using the SIS epidemic model resolves this limitation. In SIS, an infected node can still receive infections (messages) from other nodes, meaning that multiple infection paths contribute to the initiator node's influence score. The analysis shows that a strong positive correlation can be obtained between SIS and YPC, due to similarities in their parameter structures and influence mechanisms: In SIS, the recovery probability $\gamma$, which determines whether a node returns to the susceptible state and remains receptive to further infections, prevents saturation and is directly related to $\frac{k_n}{k_o}$. Additionally, Possibility of infection transmission $\beta$ is directly related to the neighbor degrees $k_n$, which is nearly identical to how $\varrho = \frac{\langle k_n \rangle}{k_o}$ in YPC directly affects influence radius and node impact. Due to this structural similarity, a positive correlation emerges. The use of $\frac{k_n}{k_o}$ in SIS provides an interpretation closely related to influence radius, as a node's ability to be reinfected means that its neighbors continue to propagate messages, leading to an extended influence spread.

However, a major challenge with SIS is the potential for infinite infection loops, where the transmission chain is never broken. This issue must be carefully managed through proper selection of $\gamma$ and $\beta$ values and computational constraints, depending on the network structure. This challenge is so significant that many researchers opt to use the SIR model instead, utilizing the recovered state (R) as a limiting factor, even when its presence lacks a strong theoretical justification for the given study.

Since in the SIR model, the recovered state (R) does not have a meaningful interpretation in the case study of this research[‡], and its inclusion would introduce unnecessary complexity in evaluating YPC, SIR is not used. However, in other network types or specific scenarios, the R state may hold relevance, making SIR a valuable model for certain applications. Considering all these factors, SIS is chosen as the primary benchmark model for evaluating YPC in this study.

### 6.2. Experimental Results and Analysis

To comprehensively evaluate the performance and generalizability of the proposed YPC model, three sets of experiments were designed to compare its node rankings with those of established centrality and epidemic-based models. The correlation between rankings was quantified using Kendall's tau coefficient. The first two experiments were performed on synthetic networks to test theoretical behavior, while the third focused on real-world social networks for empirical validation.

---

[‡] That is, if a node in a social network receives a message and becomes infected (informed), the recovered state, which is also interpreted as immunity (i.e., the inability to receive the message again), is meaningless for this node; a node can receive a message multiple times through different paths.

□ **Barabási–Albert Network**

In the first experiment, a random Barabási–Albert graph was used. The Barabási–Albert model is based on the principles of network growth and preferential attachment. Nodes are gradually added to the network and tend to connect more frequently to nodes that already have higher degrees. Structurally, this model naturally generates networks with a power-law degree distribution, where a few nodes have very high degrees while the majority have lower degrees. This model is suitable for modeling social networks, the internet, and many natural systems where highly connected nodes (hubs) exist. Figure (9) illustrates a Barabási–Albert graph consisting of 100 nodes.

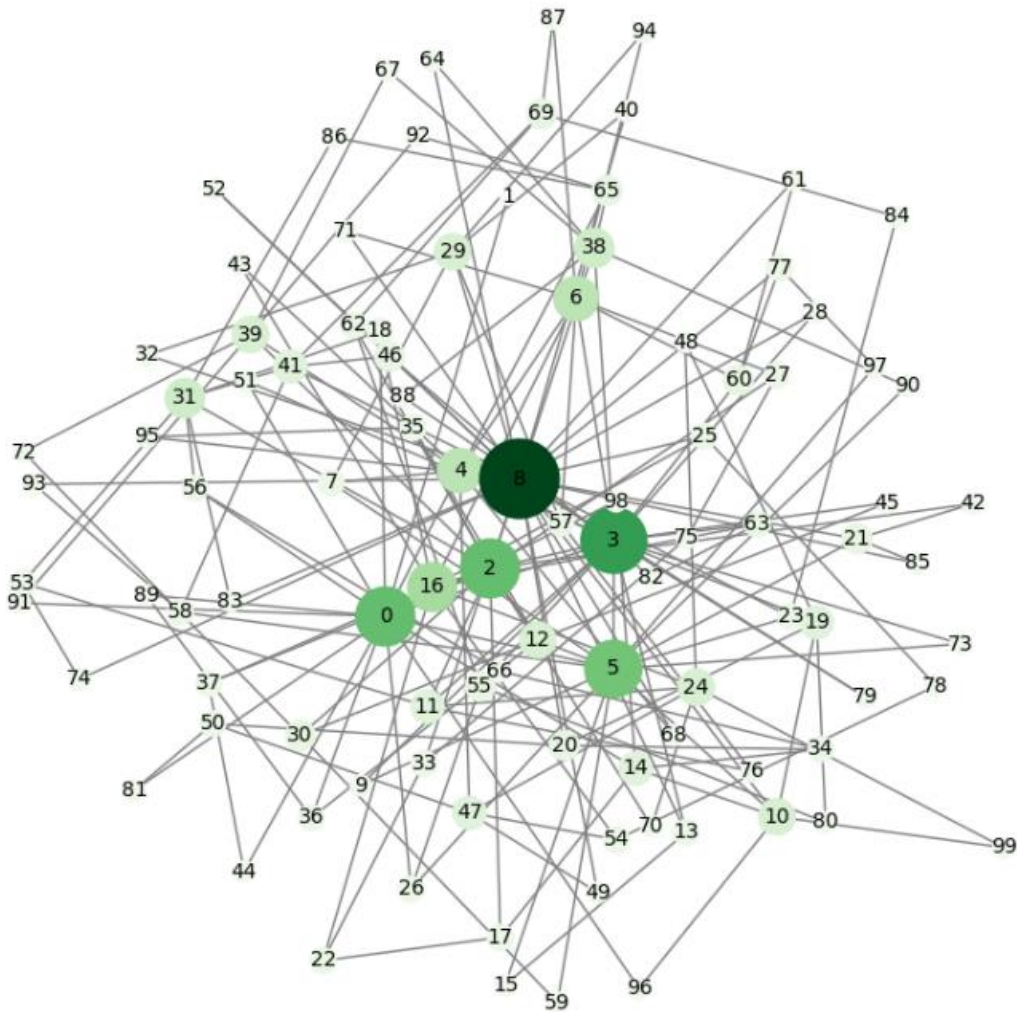

**Figure 9.** A graph of a Barabási-Albert random network with 100 nodes and 196 edges; the numbers indicate node labels. The larger and darker the circle representing a node, the higher its degree.

The Kendall's tau correlation coefficient between the ranking results of the YPC model and the SI and SIS models is presented in Table 1. As the values indicate, there is a very strong positive correlation between YPC and SI, while the correlation between YPC and SIS is moderate to strong. Given the small P-values in both cases, it can be concluded that these correlations are statistically significant, and the probability of these correlations occurring under the assumption of no association is very low. This consistency confirms that the YPC model effectively captures the diffusion dynamics observed in epidemic-based frameworks, validating its capability for realistic influence assessment.

**Table 1.** Kendall's tau correlation coefficient between the ranking results of the YPC, SI, and SIS models in the Barabási–Albert network.

|  | YPC & SI | YPC & SIS |
| --- | --- | --- |
| Kendall's Tau | 0.866787 | 0.664510 |
| P-value | $2.394803 \times 10^{-37}$ | $1.215942 \times 10^{-22}$ |

The Barabási–Albert synthetic network was selected because it exhibits a scale-free degree distribution, a property commonly observed in real social and communication networks. Evaluating YPC on this network allows assessing its capability to capture hub-dominated influence patterns.

▢ **Les Misérables Network**

The Les Misérables network is an undirected and weighted network representing the co-occurrence of characters in Victor Hugo's novel Les Misérables. This network consists of 77 nodes and 254 edges, where nodes represent characters labeled with their respective names, and edges connect any two characters who appear together in the same chapter. The edge weights (thickness) indicate the frequency of these co-occurrences. Since human relationships dominate the Les Misérables network, it has been selected for the second experiment in this study, following the case study approach. Figure (10) illustrates the Les Misérables network, where edge weights have been disregarded as per the second assumption in Section 5.1.

By applying the YPC and SIS models to the Les Misérables network and computing the Kendall's tau correlation coefficient, a moderate positive correlation can be observed, as shown in Table 2. The value of 0.4259 indicates that the ranking of nodes in both models is fairly similar, although some differences exist due to the fundamental distinctions between the two models. YPC operates based on the topological potential of nodes and a limited radius of influence, whereas SIS models message spreading more dynamically by allowing nodes to return to a susceptible state, thereby increasing their effective influence radius. Additionally, the extremely small P-value of $5.8 \times 10^{-8}$ suggests that the correlation between the two models is statistically significant, with an almost negligible probability of occurring by chance. This ensures that the results obtained are reliable and interpretable.

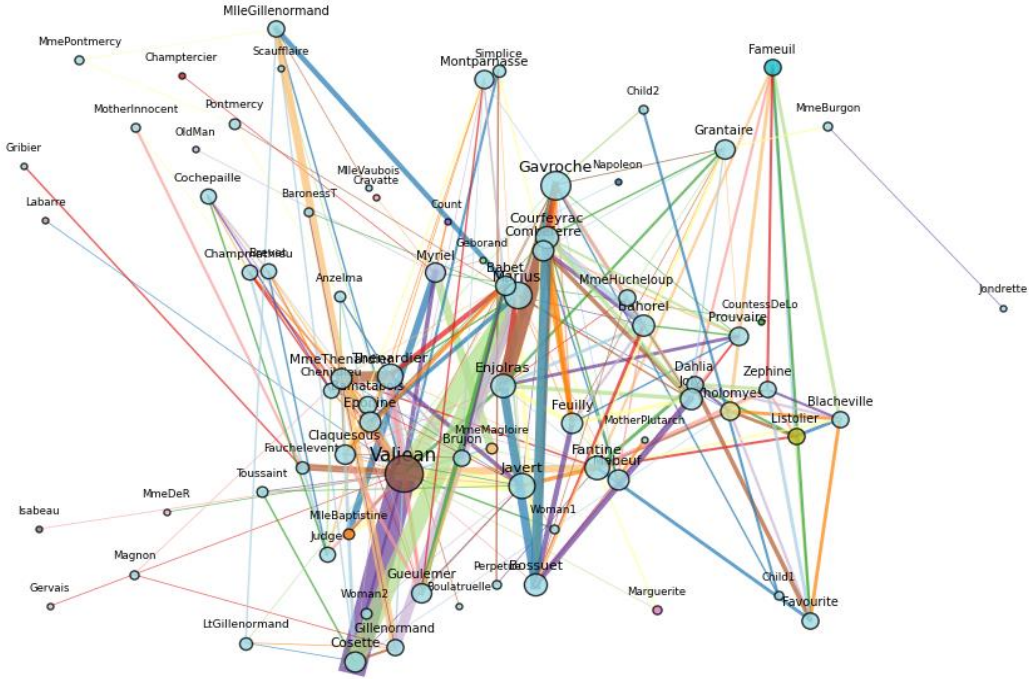

**Figure 10.** Les Misérables network. Each node represents a character.

**Table 2.** Kendall's tau correlation coefficient between the ranking results of YPC and SIS in the Les Misérables network.

|  | YPC & SIS |
|---|---|
| Kendall's Tau | 0.425949 |
| P-value | $5.799144 \times 10^{-8}$ |

The evaluation of the YPC model on the Les Misérables character network reveals that Jean Valjean, despite being a central figure in the story, receives a relatively low influence score in this metric. While this may seem contradictory at first, it can be explained by analyzing the network topology and Valjean's role in the narrative. First, due to his social limitations and lack of trust from others, Valjean does not establish an extensive social network. His isolation throughout the novel and his criminal past reduce both his number of direct neighbors and the average degree of his neighbors, thereby limiting his message-spreading capability. Second, the YPC model assesses influence within a constrained radius. Although Valjean significantly impacts characters like Cosette and Marius, this influence does not extend further into the broader network due to the model's limited radius of effect. Consequently, while Valjean plays a crucial role in the story, the YPC model strictly evaluates his topological influence within the network structure. This finding demonstrates that YPC effectively reflects the actual network structure, even when its results diverge from narrative or conceptual interpretations.

The Les Misérables network represents a small-world social structure with well-defined community semantics based on human relationships. This property provides a suitable testbed for evaluating how YPC handles socially interpretable influence propagation.

Therefore, this case demonstrates that YPC can differentiate between narrative prominence and structural influence, a valuable property when analyzing social or semantic networks.

☐ **Real-World Social Network**

In this experiment, a real-world dataset was used to evaluate the YPC model. This dataset, derived from a sample of the Facebook social network [29], consists of an undirected graph with over 4,000 nodes and 88,000 edges, where nodes represent users and edges correspond to friendship connections. The Facebook dataset was chosen because it represents a large-scale, heterogeneous, and highly clustered network, making it ideal for testing the scalability and stability of YPC on empirical data. The study of such large-scale real-world networks presents two important considerations:

1. Although the dataset used in this study originates from a reliable source, it is not entirely free of errors. For example, in the dataset file, which is described as an undirected network, it is possible that node A is recorded as being "connected" to node B, while node B does not reciprocally list a connection to node A. Such inconsistencies in the dataset can lead to computational inaccuracies.
2. In the SIS comparison model, fixed coefficients can be incorporated into the dependency equations related to the network's topological properties, as discussed in Section 6.1. These coefficients influence both computational accuracy and processing overhead. Increasing computational precision leads to higher processing costs. Therefore, determining appropriate numerical values for these coefficients is crucial to balancing result accuracy and reasonable computational overhead. With these considerations in mind and as mentioned in Section 6.1, Table 3 presents the tuning coefficients for the parameters $\beta$ and $\gamma$ in the SIS model for this experiment and the previous two experiments.

**Table 3.** Tuning coefficients for the parameters in SIS.

|  | Barabási–Albert | Les Misérables | Facebook |
|---|---|---|---|
| Parameter coefficient $\beta$ | 0.01 | 0.003 | 0.00001 |
| Parameter coefficient $\gamma$ | 0.7 | 0.7 | 0.7 |

Table 4 presents a moderate positive correlation between the ranking results obtained from the YPC and SIS models. As previously discussed, by carefully adjusting the transmission and recovery rate coefficients in the SIS model—at the cost of significantly increasing the computational overhead—it is possible to achieve an even stronger correlation. Additionally, the extremely small P-value indicates that the observed correlation is statistically significant and not due to random chance.

**Table 4.** Kendall's tau correlation coefficient between YPC and SIS ranking results in the Facebook network

|  | YPC & SIS |
|---|---|
| Kendall's Tau | 0.341377 |
| P-value | $6.564858 \times 10^{-232}$ |

The scatter plot in Figure (11) illustrates the relationship between the scores produced by the YPC and SIS models for the studied network. In this plot, the horizontal axis represents the scores computed by the YPC model, while the vertical axis shows the SIS scores. A detailed analysis of this plot, along with the linear regression line and dataset mean values, reveals several key insights: The added regression line in the scatter plot demonstrates a downward trend, indicating an inverse relationship between the scores of the two models. This means that an increase in YPC values is generally associated with a decrease in SIS values. Given the positive correlation calculated earlier, it is evident that this phenomenon arises due to fundamental differences in the criteria and formulation of these models: in the YPC model, nodes with more negative scores are considered more influential, whereas in the SIS model, nodes with higher positive scores are deemed more influential. This contrast in influence measurement criteria emphasizes the significance of the regression line, which reflects how the two models relate in assessing node influence. For example, if a node has a highly negative score in YPC, it is expected to have a correspondingly high positive score in SIS, highlighting the agreement between both models in identifying highly influential nodes.

Another notable observation is the dispersion of data points in the plot. The varying spread of points indicates that some nodes are ranked differently by the two models. These discrepancies may result from each model's sensitivity to network topology and either local or global network properties. Since the SIS model relies on transmission and recovery parameters, it may rank nodes with specific topological features differently from the YPC model. This divergence in influence definitions makes a combined analysis of both models a powerful tool for identifying key nodes and understanding network dynamics.

Furthermore, the mean SIS and YPC score lines, represented as horizontal and vertical indicators in the plot, mark the average score of each model. These lines enable an overall comparison of the two models' performance. The clustering of many points near the intersection of these lines suggests a relative agreement between the two models in certain parts of the network. However, points further from this intersection reveal significant ranking differences between the models. Overall, this pattern shows that YPC and SIS highlight overlapping but not identical sets of influential nodes, which reinforces the complementary nature of structural and dynamic approaches to influence analysis.

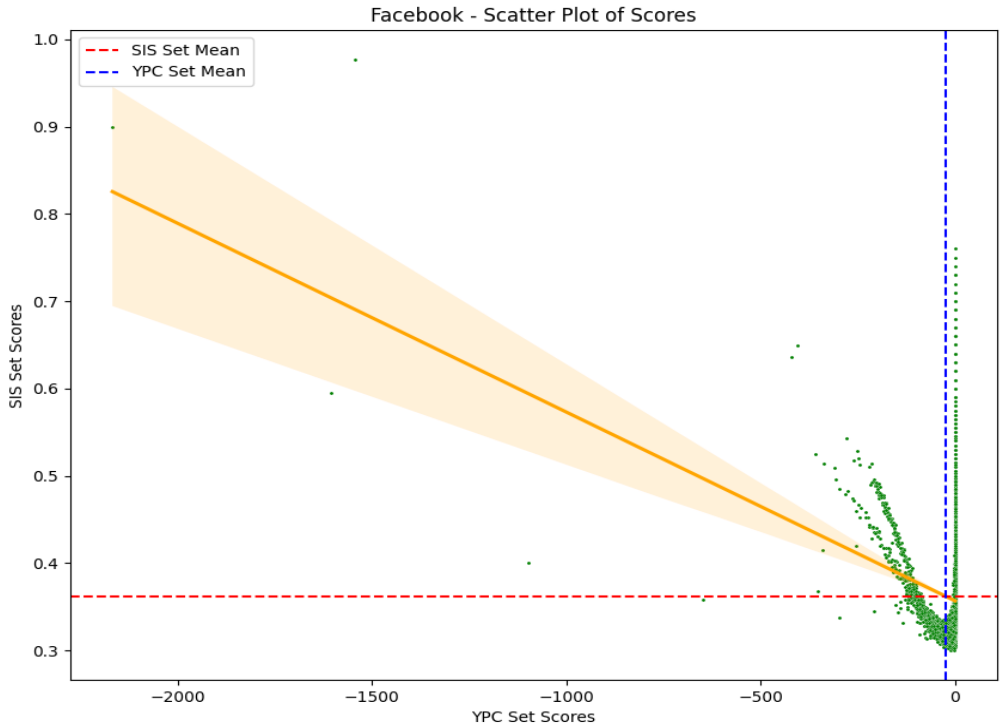

**Figure 11.**    Scatter plot of YPC and SIS scores with regression line for the Facebook network.

**Comparison with Other Centrality Measures**

Figure (12) presents ranking comparison plots between the YPC model and the centrality measures GC, DC, BC, CC, EVC, EcC, and FC for the Facebook network. In each plot, the horizontal axis represents the node index, while the vertical axis indicates its rank. The red lines correspond to the YPC model, while the blue lines represent the other centrality measures. These visual comparisons provide an intuitive understanding of how YPC aligns with or diverges from traditional metrics that capture local (e.g., degree) or global (e.g., betweenness, eigenvector) importance.

If the red and blue lines are closely aligned and follow a similar trend, this suggests a high correlation between the rankings produced by the two models. Conversely, if there is a significant gap between the lines or irregular variations in one of them, it indicates that the two models evaluate node importance differently.

These plots also demonstrate the models' sensitivity to node positions within the network. If sudden changes occur in both models at the same points, it can be inferred that they have identified some key nodes, although their final ranking may differ. Additionally, the range of rank values in each model suggests that nodes are distributed across a spectrum from high to low ranks. However, in some instances, significant discrepancies between the rankings of the two models indicate that one model considers certain nodes more influential than the other.

The plots in Figure (13) offer an alternative interpretation of node rankings in the YPC model compared to other centrality measures. In each plot, the horizontal axis represents node indices, while the vertical axis shows their ranks. Blue points indicate that YPC and the corresponding centrality measure assigned similar rankings to certain nodes, whereas red points highlight nodes where the two models have significant ranking differences. The darker the blue color, the greater the agreement between the two models for that node. Conversely, darker red points indicate greater ranking discrepancies. In this analysis, normalization was applied to unify the ranking scales across different models. Normalization ensures that ranking values are transformed within a predefined range (e.g., 0 to 1 or 0 to 100), making comparisons between models more meaningful. This adjustment eliminates the impact of inherent scale differences between ranking methods, enabling a more precise examination of agreement or divergence between models.

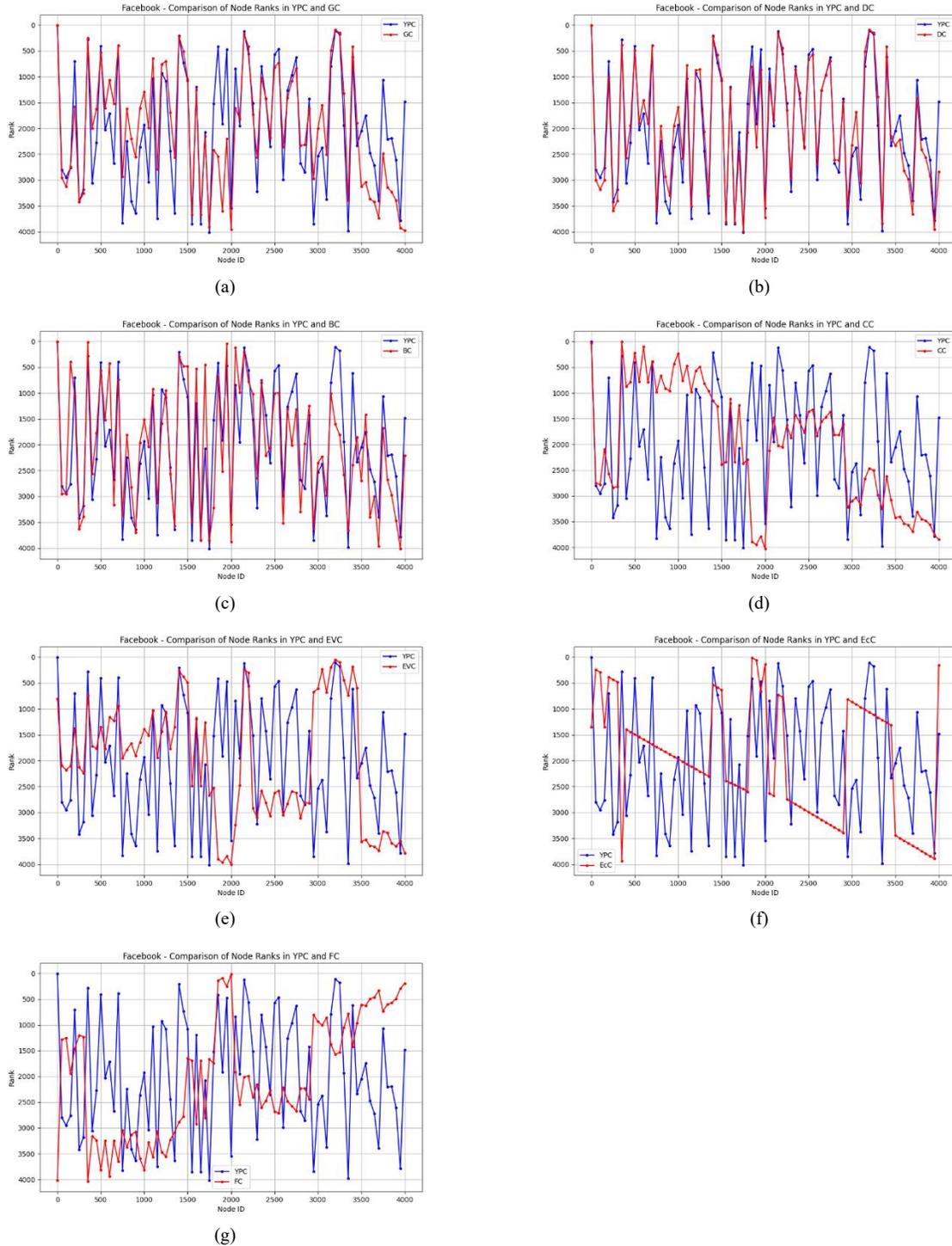

**Figure 12.** Ranking comparison plots between the YPC model and centrality measures for the Facebook network; sampling rate: every 50 nodes; (a) YPC & GC, (b) YPC & DC, (c) YPC & BC, (d) YPC & CC, (e) YPC & EVC, (f) YPC & EcC and (g) YPC & FC.

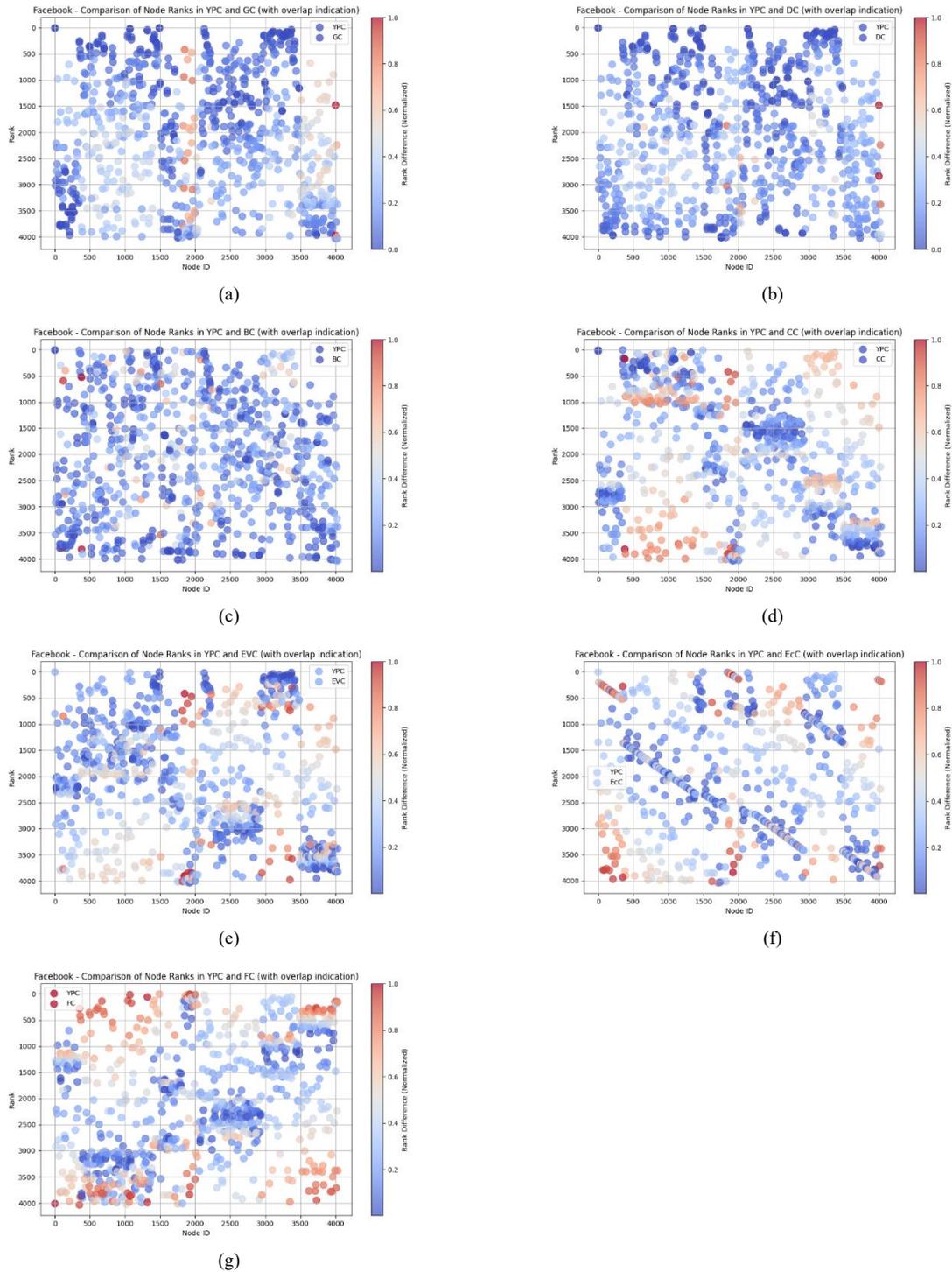

**Figure 13.** Overlap plots of node rankings in the YPC model compared to centrality measures for the Facebook network; sampling rate: every 10 nodes; (a) YPC & GC, (b) YPC & DC, (c) YPC & BC, (d) YPC & CC, (e) YPC & EVC, (f) YPC & EcC and (g) YPC & FC.

A similar experiment was conducted on the GitHub network [30], which consists of over 37,000 nodes and 289,000 edges. Nodes represent GitHub developers, with attributes such as location, starred repositories, employer, and email. Figures (14) and (15) illustrate the comparison of node rankings by YPC with other centrality measures.

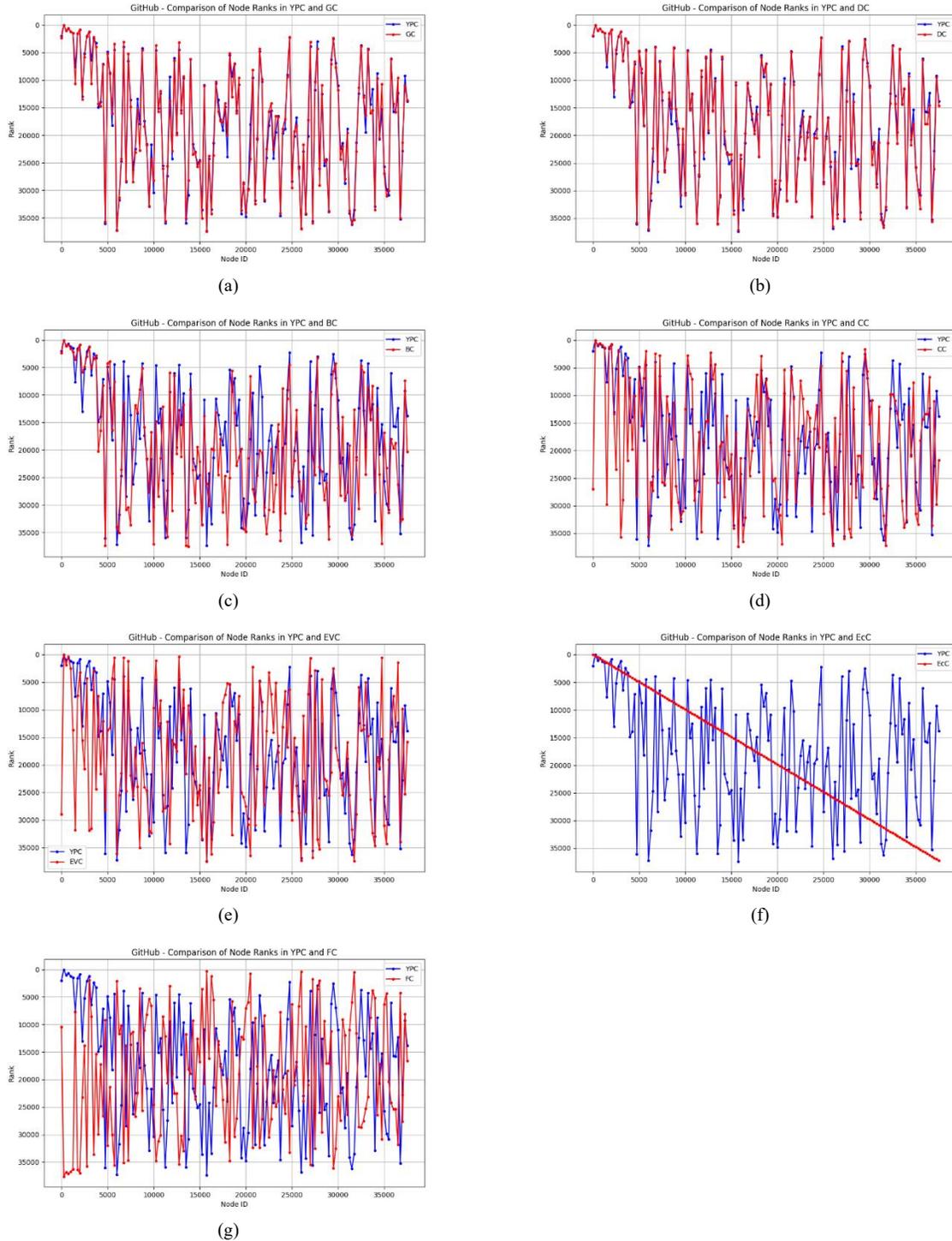

**Figure 14.** Ranking comparison plots between the YPC model and centrality measures for the GitHub network; sampling rate: every 250 nodes; (a) YPC & GC, (b) YPC & DC, (c) YPC & BC, (d) YPC & CC, (e) YPC & EVC, (f) YPC & EcC and (g) YPC & FC.

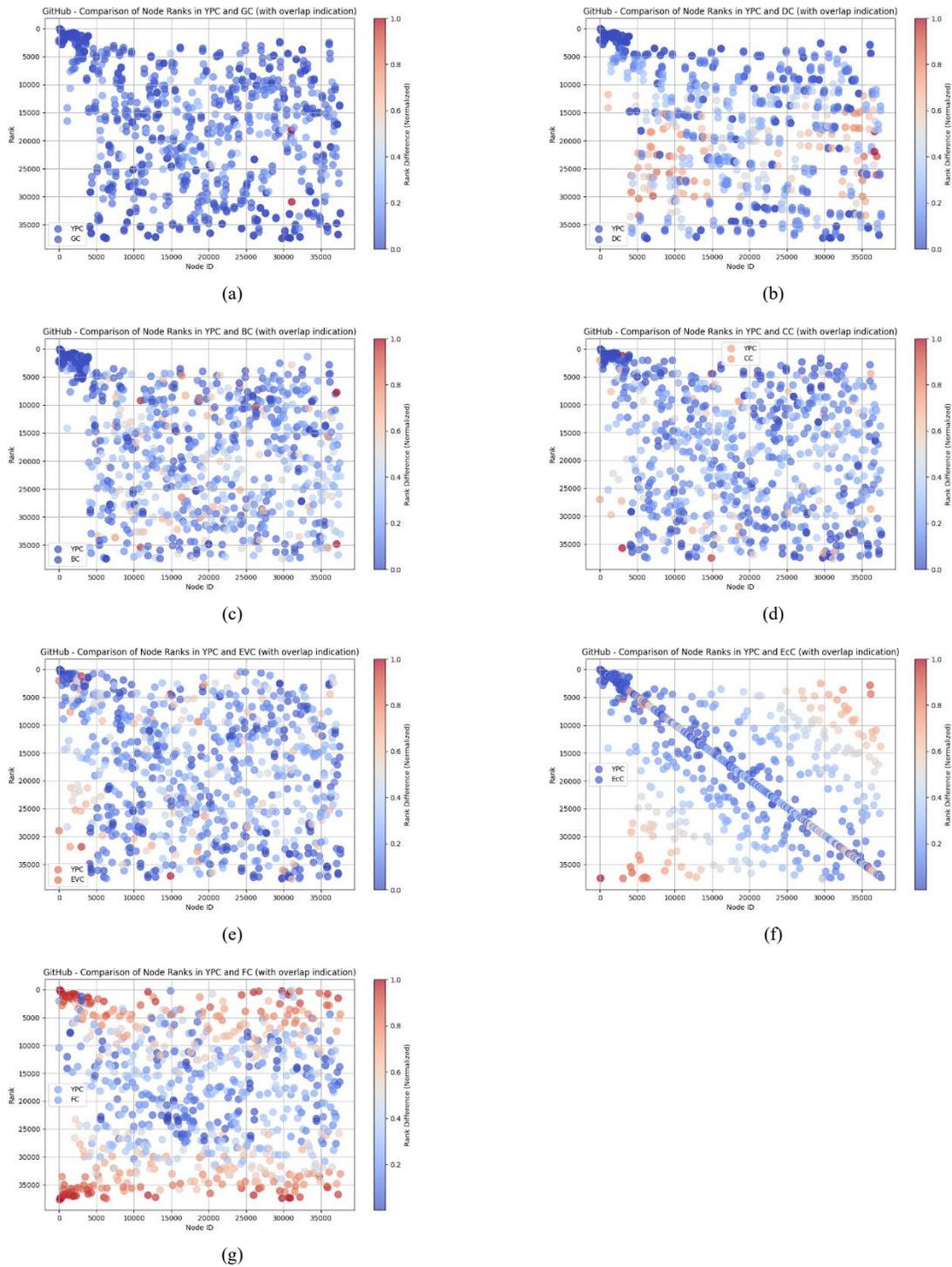

**Figure 15.** Overlap plots of node rankings in the YPC model compared to centrality measures for the GitHub network; sampling rate: every 100 nodes; (a) YPC & GC, (b) YPC & DC, (c) YPC & BC, (d) YPC & CC, (e) YPC & EVC, (f) YPC & EcC and (g) YPC & FC.

In another experiment, a network of over 7,000 nodes and 27,000 edges, representing LastFM users in Asia [31], was analyzed. The node attributes in this network are based on users' favorite artists. Figures (16) and (17) present a comparative analysis of node rankings by YPC and other centrality measures.

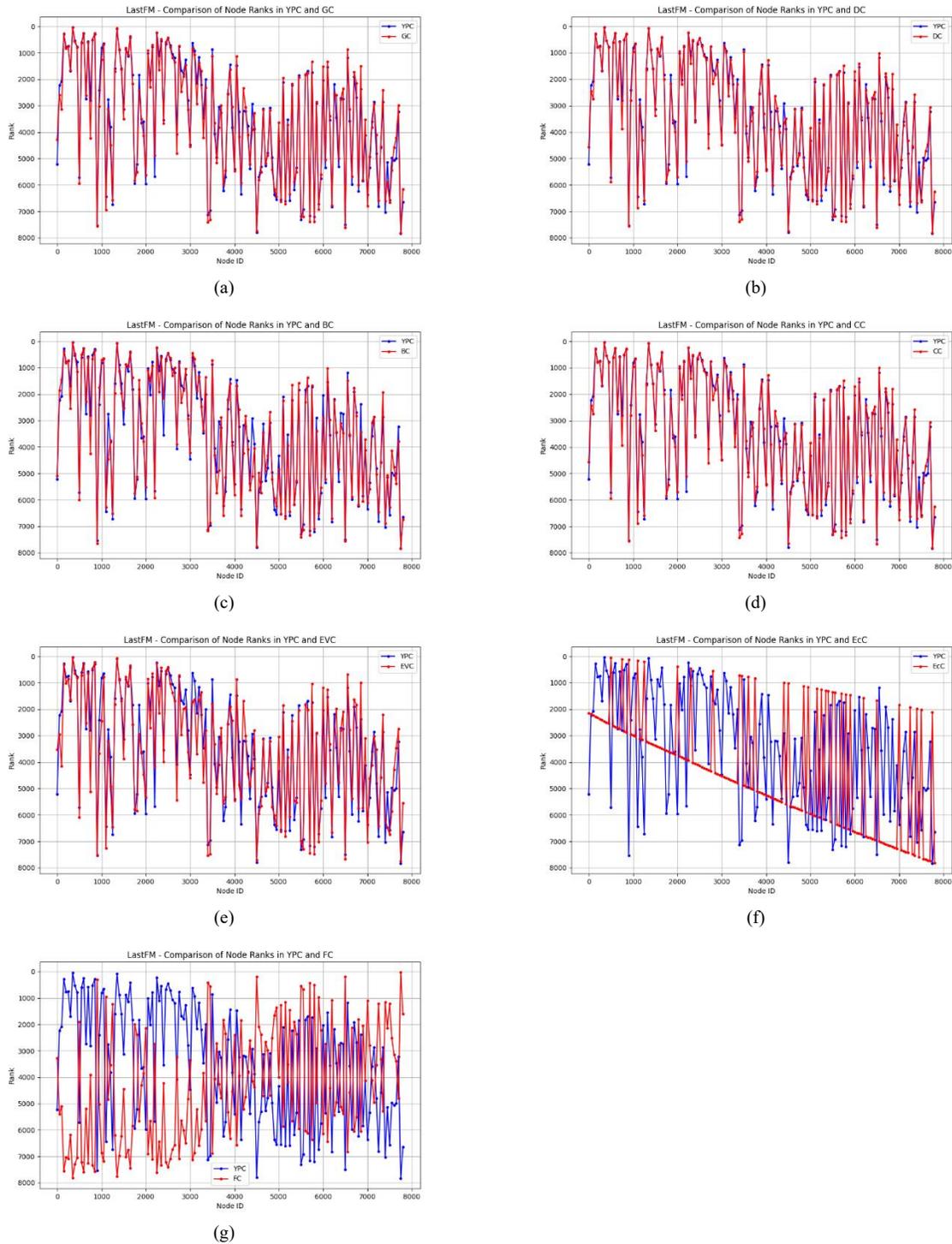

**Figure 16.** Ranking comparison plots between the YPC model and centrality measures for the LastFM network; sampling rate: every 50 nodes; (a) YPC & GC, (b) YPC & DC, (c) YPC & BC, (d) YPC & CC, (e) YPC & EVC, (f) YPC & EcC and (g) YPC & FC.

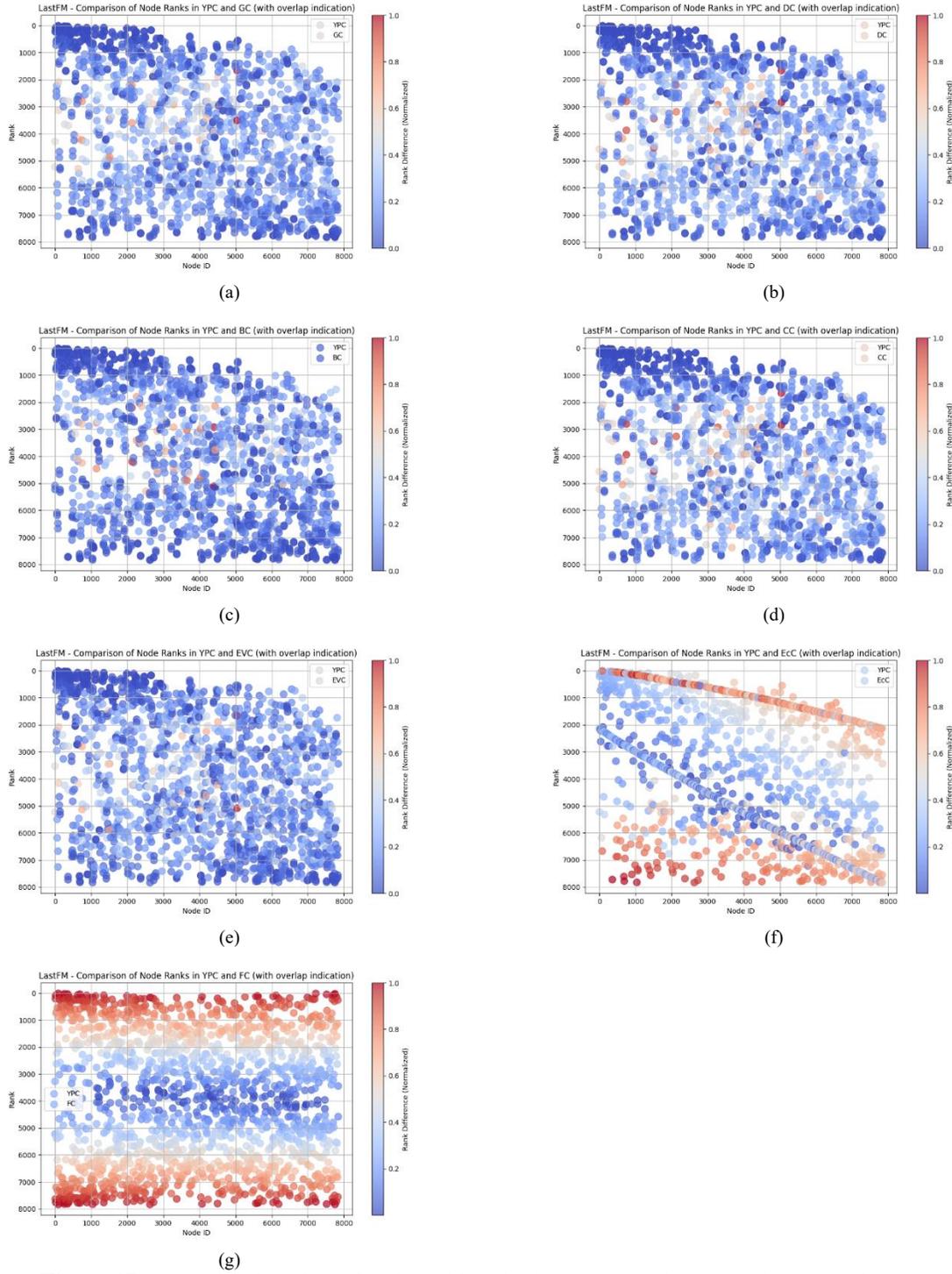

**Figure 17.** Overlap plots of node rankings in the YPC model compared to centrality measures for the LastFM network; sampling rate: every 10 nodes; (a) YPC & GC, (b) YPC & DC, (c) YPC & BC, (d) YPC & CC, (e) YPC & EVC, (f) YPC & EcC and (g) YPC & FC.

Additionally, Figures (18) and (19) illustrate the comparison of node rankings by YPC and other centrality measures in a Twitch network of Portuguese-speaking streamers [30], consisting of over 1,900 nodes and 31,000 edges. In this network, node attributes include the games played, streaming habits, and geographic location.

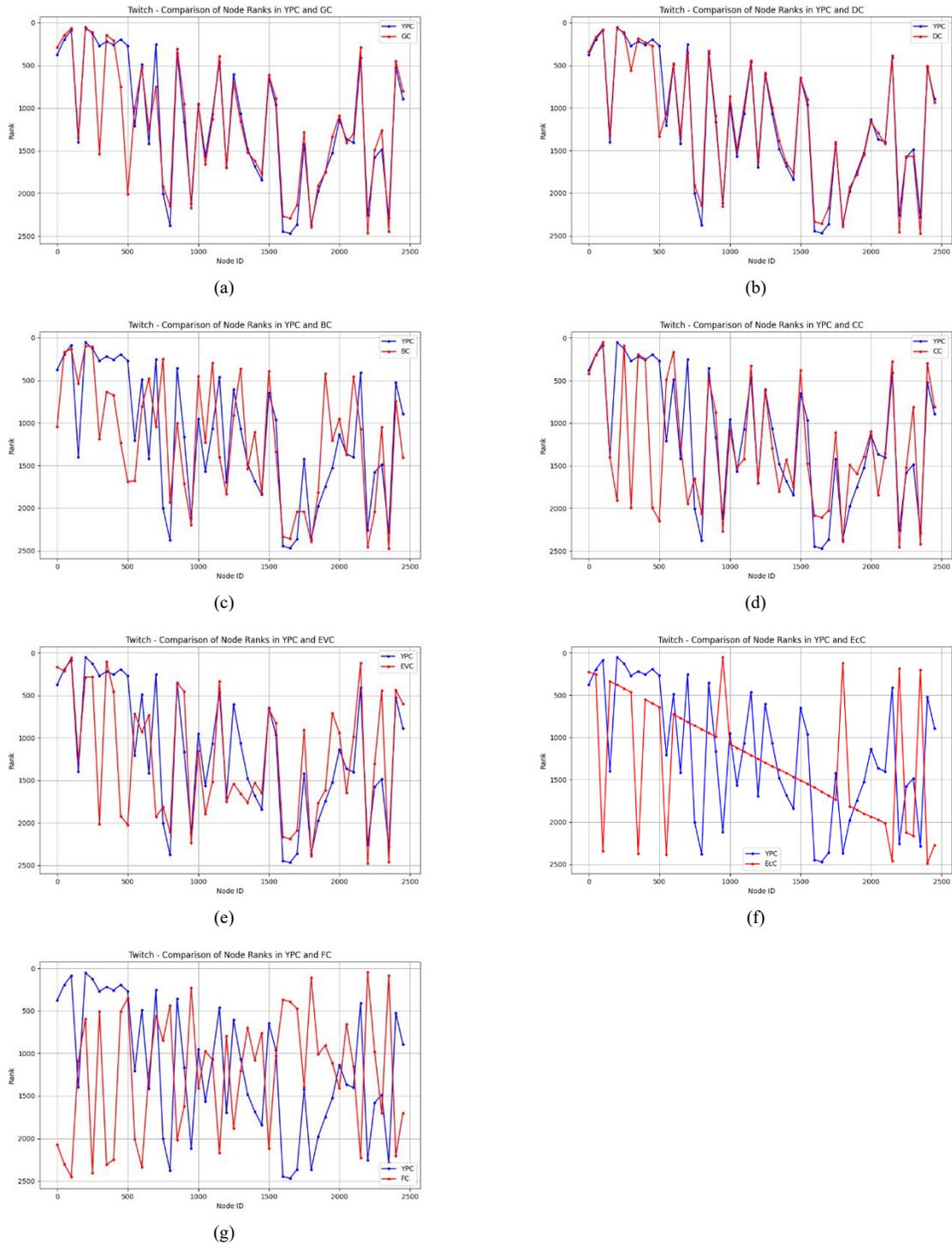

**Figure 18.** Ranking comparison plots between the YPC model and centrality measures for the Twitch network; sampling rate: every 50 nodes; (a) YPC & GC, (b) YPC & DC, (c) YPC & BC, (d) YPC & CC, (e) YPC & EVC, (f) YPC & EcC and (g) YPC & FC.

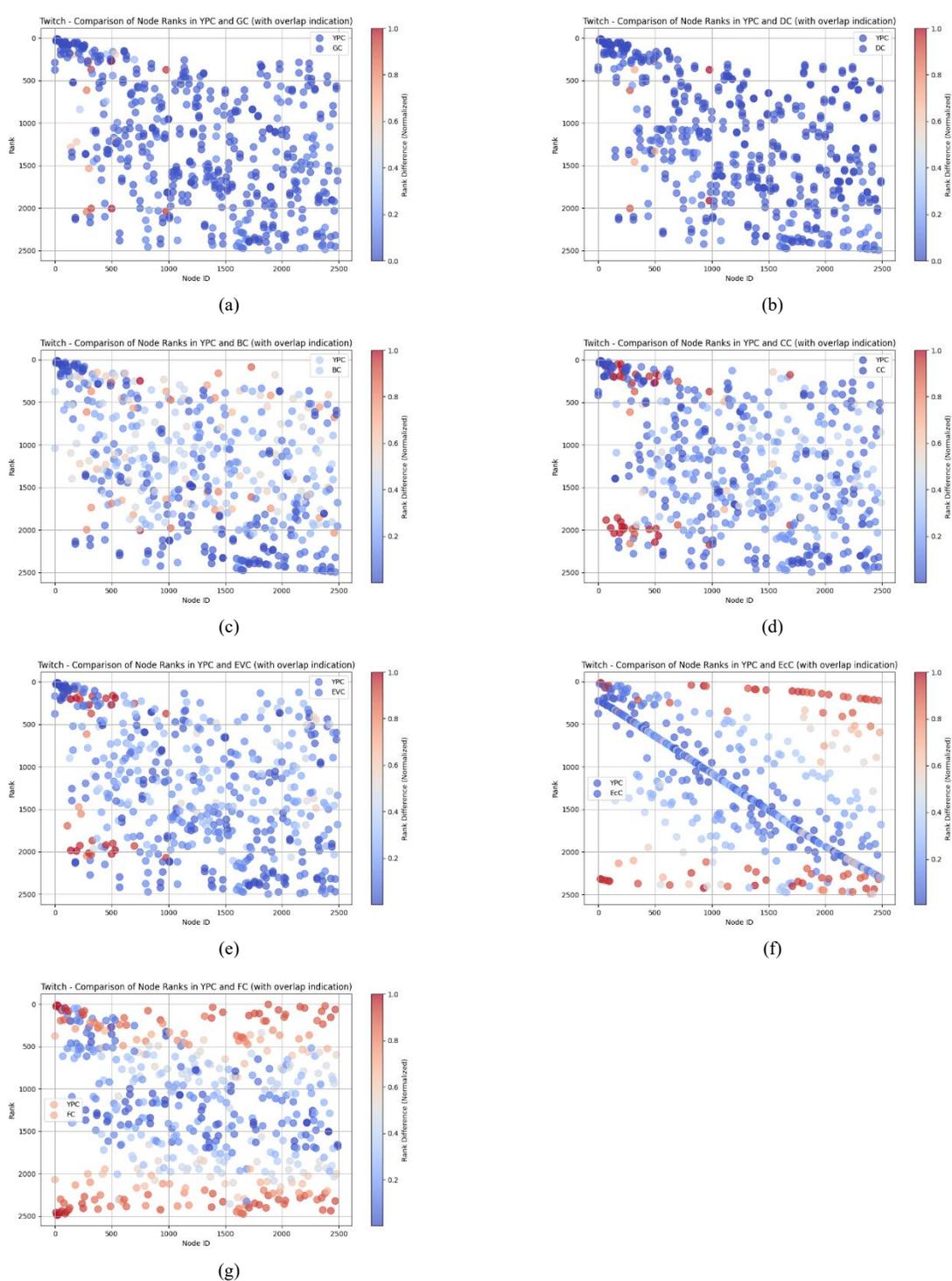

**Figure 19.** Overlap plots of node rankings in the YPC model compared to centrality measures for the Twitch network; sampling rate: every 10 nodes; (a) YPC & GC, (b) YPC & DC, (c) YPC & BC, (d) YPC & CC, (e) YPC & EVC, (f) YPC & EcC and (g) YPC & FC.

These real-world datasets with diverse interaction types (collaborative, preference-based, and behavioral) were included to evaluate YPC's adaptability across different network semantics.

**Radius of Influence**

As previously explained, in the Yukawa Potential Centrality (YPC) model, the short-range influence is determined by specifying an approximation threshold close to zero. Beyond this threshold, the calculated potential becomes negligibly small, allowing it to be disregarded. Unlike previous methods, YPC leverages the nature of the underlying physical phenomenon, causing potential values (scores) to decay sharply toward zero. This differs from classical gravity-based centrality and many other node ranking methods, where the radius of influence for message transmission is typically restricted to three neighborhood layers from each initiator node. As a result, YPC does not require an artificial cutoff radius, ensuring that computational accuracy remains intact. Figure (20) presents the radius of influence of nodes in the real-world networks analyzed. In the Facebook network, the shortest observed message spread covered four neighborhood layers, while the longest extended to eight layers, with an average radius of 6.35 hops per node. The average radius for the GitHub, LastFM, and Twitch networks was 2.99, 2.27, and 4.01 hops, respectively.

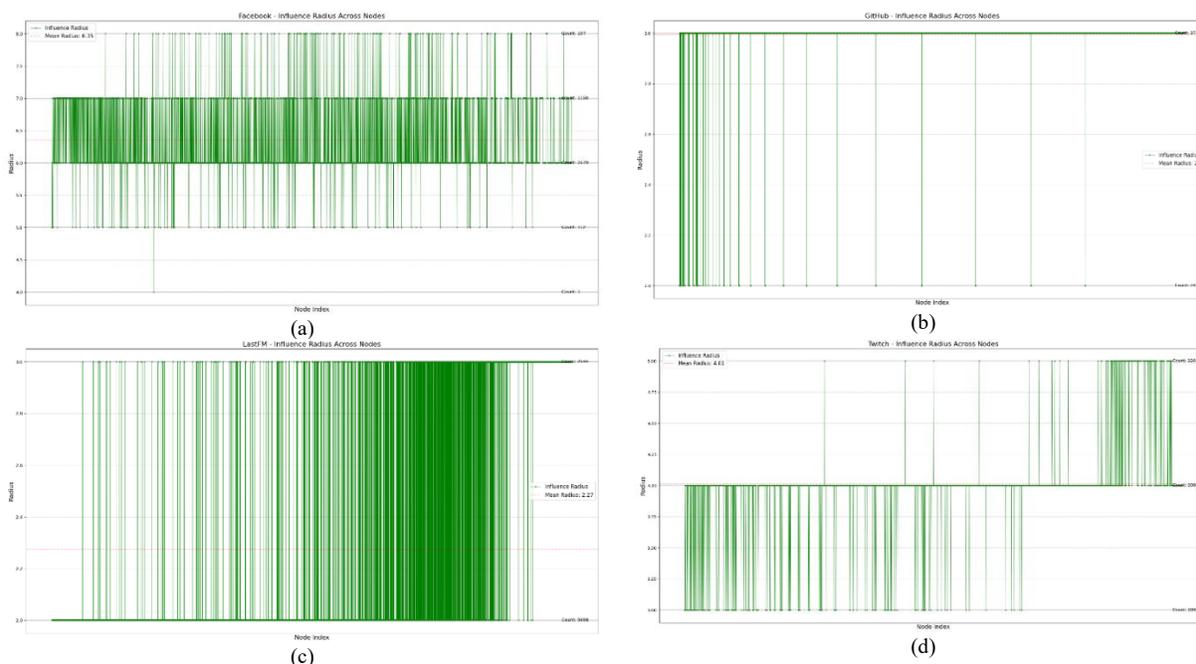

**Figure 20.** Influence radius of nodes based on the YPC model in real-world networks: (a) Facebook, (b) GitHub, (c) LastFM and (d) Twitch.

Overall, the experiments across both synthetic and real-world networks demonstrate that YPC can accurately identify key spreaders while maintaining low computational cost. Its strong statistical consistency with epidemic models and interpretability through a physical potential framework make it a promising and scalable approach for influence analysis in complex systems.

## 7. Threats to Validity

Several factors may affect the validity of the results in evaluating the Yukawa Potential Centrality (YPC) model. These threats include limitations related to the choice of studied networks, sensitivity to parameter settings, simplifying assumptions, and the structural properties of networks. The following sections examine these threats:

1. **Impact of Network Selection**
   The YPC model has been tested on various networks, including a Barabási-Albert random graph, the Les Misérables character network, and real-world networks. Each of these networks has specific topological structures that may impose limitations on the generalizability of the results. For instance, the model might perform differently in networks with high clustering coefficients or extreme heterogeneity. Therefore, further investigations are required to assess its applicability to other types of networks.
2. **Sensitivity to Parameter Settings**
   The YPC model depends on parameter settings, particularly the influence radius threshold. Variations in these parameters can lead to differences in ranking influential nodes and affect the model's performance. In some cases, improper parameter tuning may prevent the model from correctly identifying truly influential nodes. Thus, careful calibration of parameters and testing the model under diverse conditions is necessary.
3. **Impact of Model Assumptions**
   For simplicity, the study assumes that the analyzed networks are unweighted and undirected. However, in complex networks such as biological, communication, or economic networks, this assumption may overlook specific node characteristics. Additionally, YPC assumes that a node's connection to a highly influential node does not necessarily increase its own influence score. While this assumption is valid in many social networks, it may reduce accuracy in networks with strong mutual feedback mechanisms.
4. **Limitations of Influence Radius**
   The YPC model is designed based on a dynamic influence radius, meaning that a node's influence is only considered within a specific range. While this improves local influence accuracy, it may limit performance in networks where long-range influence is significant. In such cases, further optimization of the model is needed to enhance its ability to capture long-distance interactions.

## 8. Conclusions

This paper introduced the Yukawa Potential Centrality (YPC) model as a novel and physically inspired framework for identifying influential nodes in complex networks. By leveraging the mathematical properties of the Yukawa potential, the model dynamically determines node influence without requiring pairwise interaction computations. This non-interactive, scalar-potential formulation substantially reduces computational overhead while preserving analytical precision.

Experimental evaluations on both synthetic and real-world social networks demonstrated that YPC can effectively detect key nodes with high accuracy and maintain a strong statistical correlation with epidemic-based models such as SIS. These results confirm that YPC accurately reflects underlying topological structures and influence diffusion dynamics.

Unlike traditional gravity-based models that assume a fixed influence radius, YPC introduces a dynamic and adaptive influence range derived from the network topology itself. This makes the model highly flexible for analyzing heterogeneous, large-scale, and irregular networks.

Importantly, the proposed YPC model provides a bridge between physical potential theory and network science, establishing a new theoretical foundation for measuring influence propagation. It offers a scalable and interpretable approach to network analysis, paving the way for future studies that integrate physical analogies with complex system modeling.

Overall, YPC represents an innovative and computationally efficient contribution to the study of complex networks. The findings of this research open promising directions for extending influence analysis to broader domains such as social, biological, and communication systems.

## 9. Future Work

This paper presents YPC as an innovative step toward identifying influential nodes in complex networks. However, several research directions can enhance and expand its capabilities:

- **Maximizing Influence Propagation via Field Theory**
  By leveraging the field equations derived from the Yukawa potential, future studies can optimize influence pathways in networks. For example, strategically placing benchmark nodes within a network could optimize message diffusion and identify the most influential regions. This approach could significantly improve information flow and resource allocation in complex networks.
- **Applying YPC to Directed and Weighted Graphs**
  The current study evaluates YPC on undirected and unweighted networks. Extending the model to directed and weighted graphs can provide deeper insights into node behavior in more complex network structures. In directed networks, one-way interactions play a crucial role, while in weighted networks, the strength of connections can significantly impact node influence. Investigating these aspects could improve YPC's accuracy and adaptability.
- **Expanding Case Studies**
  Current evaluations of YPC are mainly focused on social networks. Extending its application to biological, transportation, and financial networks could assess its generalizability across diverse domains. Additionally, exploring new topological mappings for the Yukawa potential equation parameters may provide fresh analytical perspectives for future studies.

These research directions will contribute to advancing the YPC model and expanding its applications in complex network analysis. Future studies should explore YPC's performance in real-world networks with diverse structures and datasets to further refine its capabilities.